\documentclass[aps,pre,showpacs,twocolumn]{revtex4}
\usepackage{bm}
\usepackage{graphicx}
\bibstyle{approve.bib}
\usepackage{amssymb}
\usepackage{amsmath}
\usepackage{esint}
\usepackage{epstopdf}
\usepackage{color}
\usepackage{gensymb}
\usepackage{mathtools}
\usepackage{suffix}

\newcommand{\be}{\begin{equation}}
\newcommand{\ee}{\end{equation}}
\newcommand{\bea}{\begin{eqnarray}}
\newcommand{\eea}{\end{eqnarray}}
\newcommand{\lan}{\left\langle}
\newcommand{\ran}{\right\rangle}
\newcommand{\br}{\mathbf{r}}

\newcommand{\bq}{\mathbf{q}}

\newcommand{\bk}{\mathbf{k}}

\newcommand{\bl}{\mathbf{l}}

\newcommand{\e}{\varepsilon}

\newcommand{\tv}{\tilde{v}}

\newcommand{\td}{\tilde{\Delta}}

\newcommand{\tz}{\tilde{z}}
\newcommand{\tL}{\tilde{L}}

\newcommand{\bt}{\bar{\tau}}

\newcommand{\pa}{\parallel}

\newcommand{\zm}{\tilde{z}_-}
\newcommand{\zp}{\tilde{z}_+}

\newcommand{\s}{{\rm s}}

\newcommand{\ct}{\cos\te}
\newcommand{\st}{\sin\te}
\newcommand{\te}{\theta_{\rm p}}
\newcommand{\vp}{\varphi_{\rm p}}
\newcommand{\cpk}{\cos\phi_k}

\newcommand{\de}{\partial_z}
\newcommand{\md}{\mathrm{d}}
\newcommand{\mb}{\mathrm{B}}
\newcommand{\p}{{\rm p}}
\newcommand{\qc}{q_{\rm c}}

\newcommand{\ew}{\varepsilon_{\rm w}}
\newcommand{\bph}{\bar{\phi}}
\newcommand{\C}{{\rm c}}
\newcommand{\lt}{\tilde{l}_\theta(\tz)}
\newcommand{\RP}[1]{\textcolor{black} {#1}}
\newcommand{\SB}[1]{\textcolor{black} {#1}}

\begin{document}

\title{\RP{Like-charge polymer-membrane complexation mediated by multivalent cations:\\ one-loop-dressed strong coupling theory}}

\author{Sahin Buyukdagli$^{1}$\footnote{email:~\texttt{buyukdagli@fen.bilkent.edu.tr}}  
and Rudolf Podgornik$^{2,3,4}$\footnote{email:~\texttt{podgornikrudolf@ucas.ac.cn}}}
\address{$^1$Department of Physics, Bilkent University, Ankara 06800, Turkey\\
$^2$School of Physical Sciences and Kavli Institute for Theoretical Sciences,
University of Chinese Academy of Sciences, Beijing 100049, China\\
$^3$CAS Key Laboratory of Soft Matter Physics, Institute of Physics,
Chinese Academy of Sciences (CAS), Beijing 100190, China\\
$^4$Department of Physics, Faculty of Mathematics and Physics, University of Ljubljana,
and Department of Theoretical Physics, J. Stefan Institute, 1000 Ljubljana, Slovenia}

\begin{abstract}
We probe the electrostatic mechanism driving adsorption of polyelectrolytes onto like-charged membranes upon the addition of tri- and \RP{tetravalent} counterions to a \RP{bathing} monovalent salt solution. We develop a \textit{one-loop-dressed strong coupling theory} that treats the monovalent salt at the electrostatic one-loop level and the multivalent counterions within a strong-coupling approach. It is shown that the adhesive force of the multivalent counterions mediating the like-charge adsorption arises from their strong condensation at the charged membrane. The resulting interfacial counterion excess locally maximizes the screening ability of the electrolyte and minimizes the electrostatic polymer grand potential. This translates into an attractive force that \RP{pulls} the polymer to  the  similarly charged membrane. We show that the high counterion valency enables this adsorption transition even at weakly charged membranes. \RP{Additionally, strongly} charged membranes give rise to salt-induced correlations and intensify the interfacial multivalent counterion condensation, \RP{strenghtening} the complexation of the polymer with the like-charged membrane, \RP{as well as triggering} the orientational transition of the molecule prior to its adsorption. Finally, our theory provides two additional key features \RP{as evidenced} by previous adsorption experiments: \RP{first}, the critical counterion concentration for polymer adsorption decreases with the rise of the counterion valency, \RP{and second}, the addition of monovalent salt enhances the screening of the membrane charges and suppresses salt correlations. This weakens the interfacial multivalent counterion condensation and results in the desorption of the polymer from the substrate. 
\end{abstract}

\pacs{05.20.Jj,82.45.Gj,82.35.Rs}

\date{\today}
\maketitle   

\section{Introduction}

In biological systems, exotic electrostatic phenomena challenging our intuition emerge \RP{consistently} from the presence of multivalent charges ~\cite{biomatter,Naji2010}. From the electrophoretic drag of anionic polymers along the applied electric field~\cite{Aksimentiev2010,Qiu2015,Buy2018} to the folding of strongly charged biopolymers~\cite{Golestanian1999,Muthu2004,Baigl2005,Claessen2008} or condensation of like-charged polyelectrolyte solutions~\cite{Delsanti1994,Ha1997,Podgornik1998,Raspaud1999,Shkl1999,Sabbagh2000,Solis2000}, a large variety of unconventional electrostatic effects have been so far observed in diverse systems whose common characteristics is the presence of multivalent charges. Naturally, this universality has motivated intensive scientific \RP{endeavour} in order to identify the nature of the seemingly counterintuitive forces mediated by multivalent ions.

\begin{figure}
\includegraphics[width=1.\linewidth]{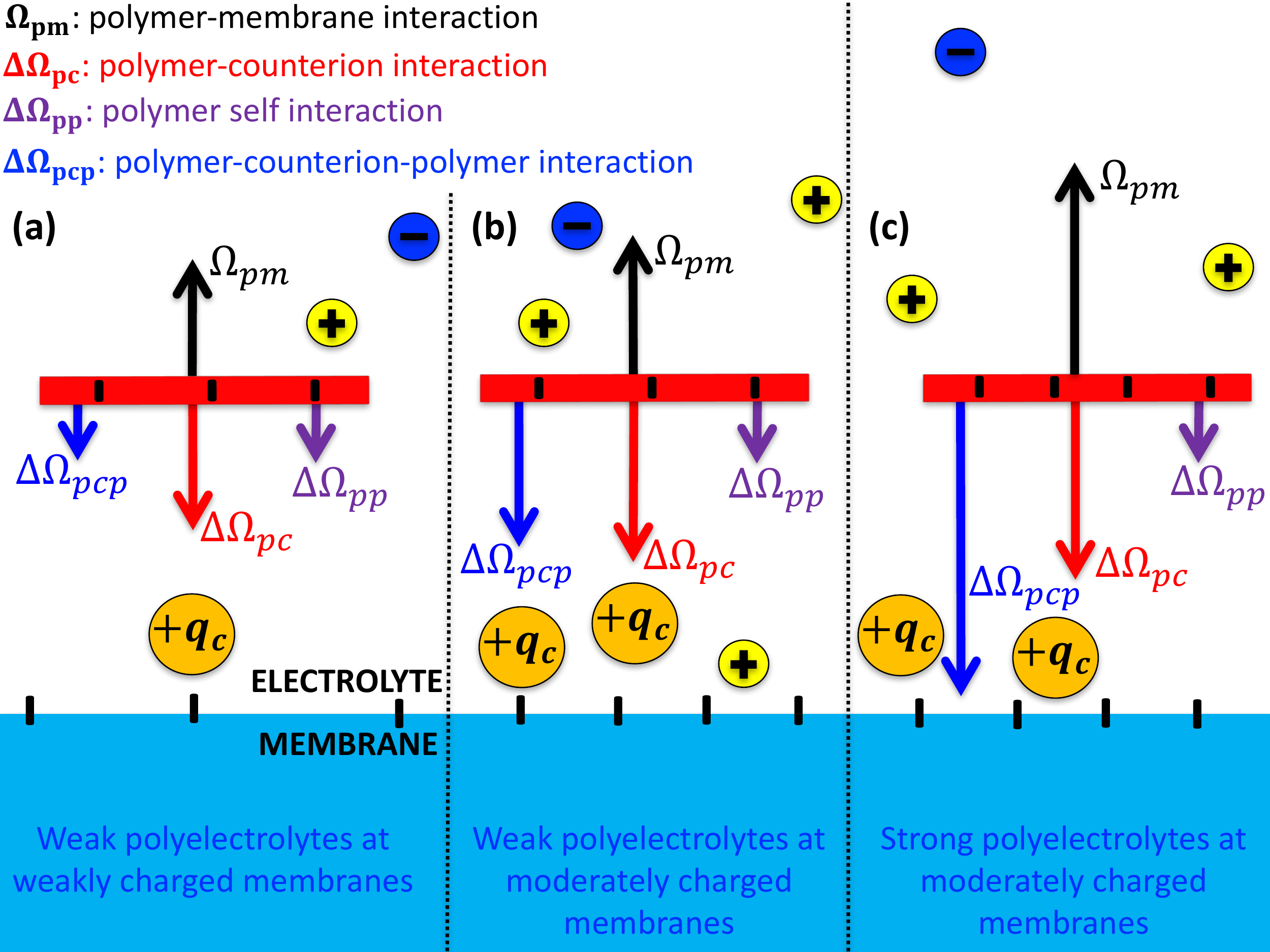}
\caption{(Color online) Schematic depiction of the polymer-liquid-membrane interactions at different polymer and membrane charge magnitudes. The anionic polymer (red) is immersed in a charged liquid composed of monovalent salt ions (blue and yellow) and multivalent counterions of valency $q_{\rm c}$ (orange). The arrows of different color indicate the magnitude of the interaction potential components in Eq.~(\ref{EqInt}).}
\label{fig0}
\end{figure}

The characterization of the effects triggered by multivalent ions requires a theoretical framework able to handle the strong-coupling (SC) \RP{electrostatic} interactions induced by their \RP{elevated} charge. A systematic perturbative theory of SC electrostatics has been developed for counterion liquids by Moretira and Netz in Ref.~\cite{NetzSC}. \RP{However, one should note} that the peculiarity of biological systems is the omnipresence of monovalent salt ions. Thus, the counterion-only formalism of Ref.~\cite{NetzSC} has been subsequently generalized by Kandu\v c et al. to the case of mixed electrolytes composed of monovalent salt and multivalent counterions~\cite{Podgornik2010}. The corresponding \textit{dressed ion theory} has been used to understand the image charge effects~\cite{Podgornik2011}, the charge regulation effects in macromolecular interactions~\cite{Adzic2016} as well as the alteration of DLVO forces by multivalent charges~\cite{Kanduc2017}.

The SC formalism of Ref.~\cite{Podgornik2010} treated the multivalent counterions within the SC approach equivalent to a low fugacity expansion while the monovalent salt was handled at the linear Debye-H\"{u}ckel (DH) level. In the present work, we upgrade this formalism by including a higher order loop correction. Namely, we develop a one-loop (\RP{$1\ell$})-dressed strong coupling theory where the multivalent counterions are considered at the SC-level but the charge fluctuations of the background salt are treated at the non-linear \RP{$1\ell$}-level~\cite{Buyuk2012}. Within this formalism, we investigate the electrostatic mechanism behind the experimentally observed polyelectrolyte adsorption onto like-charged membranes upon the addition of multivalent counterions into a monovalent salt solution~\cite{Caball2014,Tiraferri2015,Fries2017}. 

The understanding of the like-charge polymer attraction mechanism is essential for improving our control over various molecular manipulation techniques such as DNA sequencing by polymer translocation~\cite{Qiu2015} or gene delivery by DNA-liposome complexation~\cite{PodgornikRev}. It should be noted that the common gene transfer techniques are based on the use of DNA-cationic liposome complexes of high toxicity and weak biocompatibility in the cell medium~\cite{Molina2014II}. Consequently, the use of DNA-anionic liposomes of lower toxicity presents itself as a more efficient approach for gene transfer. However, the electrostatic stability of such complexes is known to occur typically under the effect of multivalent counterions. This is the point where the accurate characterization of the adhesive forces generated by multivalent cations becomes crucial.

Like-charge polymer-membrane complexation has been previously investigated by weak to intermediate coupling theories that take into account electrostatic correlations at the level of gaussian fluctuations around the mean-field (MF) Poisson-Boltzmann (PB) electrostatics~\cite{Sens2000,Buyuk2016,PRE2019}. It is known that such a gaussian closure is not adequate for the modeling of strong-coupling interactions mediated by counterions of high valency~\cite{Buyuk2014}. Motivated by this point, we develop herein the first theoretical attempt to overcome this limitation via the inclusion of SC electrostatics. 

In Sec.~\ref{th}, we present the polyelectrolyte model and review the test-charge approach of Ref.~\cite{PRE2019} previously introduced at the pure \RP{$1\ell$}-level. Then, we develop the \RP{$1\ell$}-dressed SC theory that allows to extend the formalism of Ref.~\cite{PRE2019} to the presence of strongly interacting multivalent counterions. Within this formalism, we find that the electrostatic polymer grand potential $\Delta\Omega_{\rm p}$ characterizing polymer-membrane interactions is composed of four components,
\be
\label{EqInt}
\Delta\Omega_{\rm p}=\Omega_{\rm pm}+\Delta\Omega_{\rm pc}+\Delta\Omega_{\rm pp}+\Delta\Omega_{\rm pcp}.
\ee
Fig.~\ref{fig0} illustrates the relative weight of the potential components and the charge composition of the system. The first term on the r.h.s. of Eq.~(\ref{EqInt}) corresponds to the  repulsive polymer-membrane charge coupling energy $\Omega_{\rm pm}$. The second attractive term $\Delta\Omega_{\rm pc}$ originates from the interactions of the polymer with the multivalent counterions condensed at the interface. Then, the attractive potential $\Delta\Omega_{\rm pp}$ is the polymer self-energy. Being independent of the multivalent counterions, the self-energy plays a perturbative role at all charge magnitudes considered in this work. Finally, the energy component $\Delta\Omega_{\rm pcp}$ of attractive nature accounts for the screening of the polymer self-interaction  by the interfacial multivalent counterions. 

In Sec.~\ref{mfsalt}, we consider the interaction of a weakly charged polymer with a membrane. Fig.~\ref{fig0}(a) shows that in this regime, like-charge polymer-membrane attraction is governed by the competition between the polymer-membrane interaction energy $\Omega_{\rm pm}$ and the polymer-counterion coupling potential $\Delta\Omega_{\rm pc}$. Then, Sec.~\ref{shpol} focuses on the case of intermediate membrane charges. As illustrated in Fig.~\ref{fig0}(b), we find that the increment of the membrane charge beyond the weak-coupling (WC) regime results in the emergence of monovalent salt correlations and intensifies the multivalent counterion excess. This amplifies the attractive potentials $\Delta\Omega_{\rm pc}$ and $\Delta\Omega_{\rm pcp}$, strengthens the like-charge polymer attraction, and also results in the orientational transition of the polymer from parallel to perpendicular configuration prior to its adsorption by the membrane. In addition, we show that our formalism can reproduce and explain two key features observed in previous adsorption experiments~\cite{Tiraferri2015}. First, via the enhancement of charge correlations, the increase of the counterion valency lowers the minimum multivalent cation density for the occurrence of the like-charge adsorption. Second, the increment of the monovalent salt concentration suppresses charge correlations and results in the desorption of the polymer from the membrane.

Finally, in Sec.~\ref{lgpol}, we focus on the case of strongly anionic polyelectrolytes where the self-interaction screening energy $\Delta\Omega_{\rm pcp}$ becomes the dominant attractive potential component (see Fig.~\ref{fig0}(c)). This indicates that the adsorption of strongly charged polymers such as DNA molecules is driven by the interplay between the screening energy $\Delta\Omega_{\rm pcp}$ and the repulsive polymer-membrane coupling energy $\Omega_{\rm pm}$. The limitations of our model and possible improvements are discussed in Conclusions.

\section{Theory}
\label{th}

\subsection{Polymer-membrane model}

\begin{figure}
\includegraphics[width=1.\linewidth]{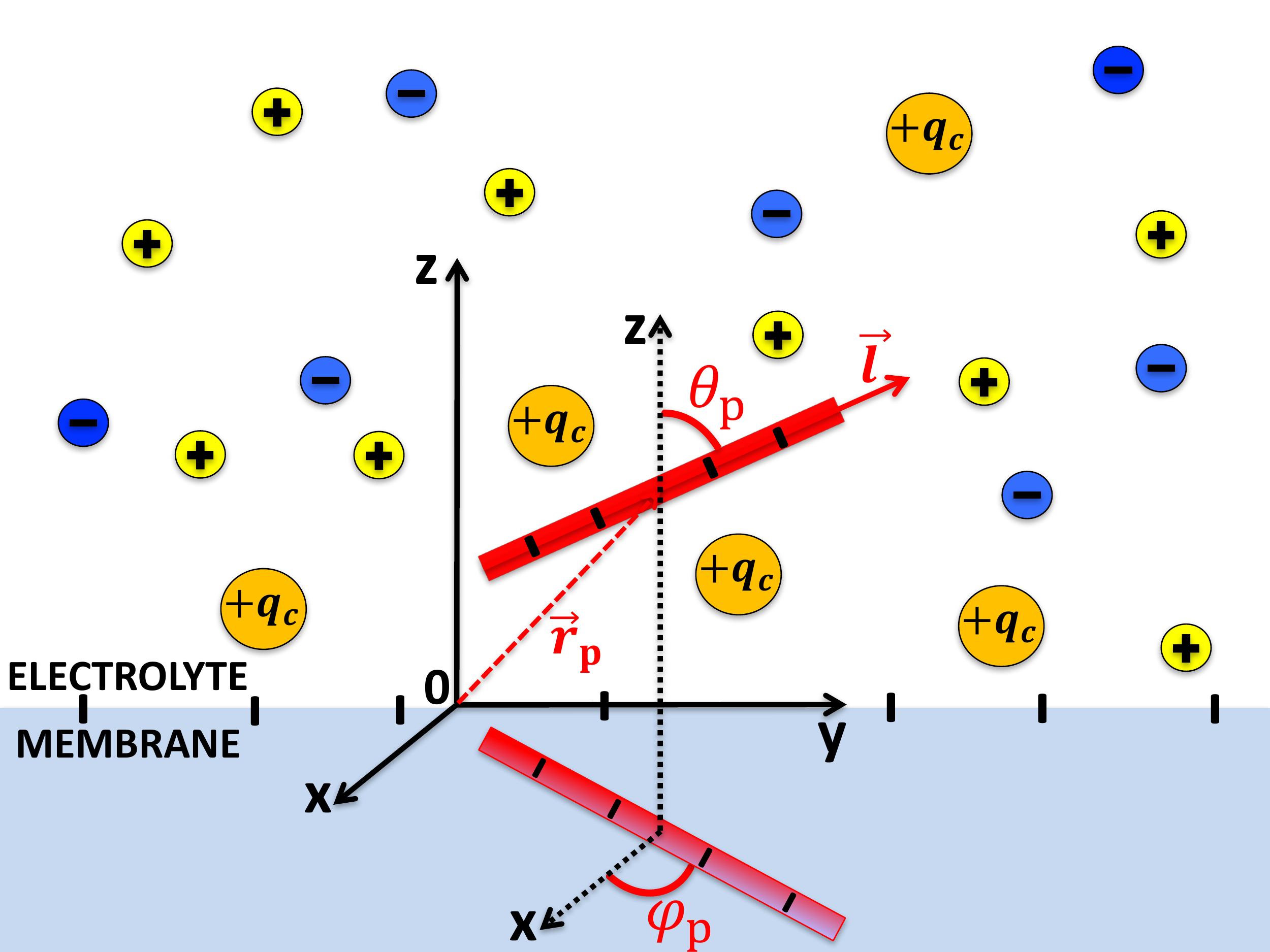}
\caption{(Color online) Depiction of the polymer-membrane complex. The polymer rotations are described by the spherical angles $\theta_\p$ and $\varphi_\p$. The length of the corotating coordinate $l$ on the polyelectrolyte is defined in the interval $-L/2\leq l\leq L/2$ where $L$ stands for the polymer length. The polymer position vector $\br_\p=(x_\p,y_\p,z_\p)$ points the geometric center of the molecule located at $l=0$.}
\label{fig1}
\end{figure}

We introduce here the charge composition of the polymer-membrane complex.  The charged system is depicted in Fig~\ref{fig1}.  The planar membrane \RP{is assumed to occupy the half-space $z\leq0$, and carries a negative surface charge density located at its surface located at $z = 0$}, 
\be\label{sigm}
\sigma_{\rm m}(\br)=-\sigma_{\rm m}\delta(z).
\ee
We neglect dielectric discontinuities and \RP{thus do not delve into the dielectric image effects, taking} $\e(\br)=\e(z)=\e_0\e_{\rm w}$ where \SB{$\e_0$ is the dielectric permittivity of vacuum and $\e_{\rm w}=80$ is the relative permittivity} of the electrolyte solution located at $z\geq0$. The electrolyte is composed of monovalent cations and anions with fugacities $\Lambda_\pm$ and bulk density $\rho_{\rm b}$, \RP{while the} multivalent counterion species has fugacity $\Lambda_\C$, valency $q_\C$, and bulk concentration $\rho_{\rm bc}$. The temperature of the electrolyte solution is $T=300$ K. 

The anionic polyelectrolyte is a stiff rod of length $L$ and linear charge density $-\tau$.  The stiff polymer approximation is motivated by the fact that the polymer length $L=5$ nm considered in this work is an order of magnitude shorter than the persistance length $\ell_{\rm p}=50$ nm of DNA molecules. Moreover, in this article, the numerical value of the polymer charge density will be expressed in terms of the double stranded DNA (dsDNA) charge $\tau_{\rm DNA}$, with the dimensionless charge density $\bt$ defined as
\be
\label{taured}
\bt=\frac{\tau}{\tau_{\rm DNA}};\hspace{1cm}\tau_{\rm DNA}=\frac{2}{3.4}\;\mbox{{\AA}}^{-1}.
\ee
The orientation of the polymer with the center-of-mass (CM) coordinate  $\br_\p=(x_\p,y_\p,z_\p)$ will be described by the azimuthal and polar angles $\te$ and $\vp$. Furthermore, we will express the polymer charge distribution in terms of the corotating coordinate $\bl$ whose magnitude is defined in the interval $-L/2\leq l\leq L/2$. The corotating coordinate system allows to express the Cartesian coordinates on the polymer in the parametric form
\bea
\label{c1}
x(l)&=&x_\p+l\st\cos\varphi_\p,\\
\label{c2}
y(l)&=&y_\p+l\st\sin\varphi_\p,\\
\label{c3}
z(l)&=&z_\p+l\ct.
\eea
Taking \SB{now} into account the impenetrability of the membrane by the polymer edges, i.e. $z(l=\pm L/2)\geq0$, one finds that the polymer rotations are limited to the interval $\theta_-\leq\te\leq\theta_+$ with the cut-off angles
\be\label{thmn}
\theta_-=\arccos\left\{\mathrm{min}\left(1,\frac{2z_\p}{L}\right)\right\},\hspace{3mm}\theta_+=\pi-\theta_-.
\ee

\SB{We finally emphasize that in this work, the interaction energies and electrostatic potentials will be expressed in dimensionless form. More precisely, the dimensionless energies will be defined as their physical counterpart rescaled by the thermal energy $k_{\rm B}T$, with the Boltzmann constant $k_{\rm B}$. Moreover, the dimensionless electrostatic potential $\phi(\br)$ will be defined in terms of the physical potential $V(\br)$ as $\phi(\br)=\beta eV(\br)$, with the inverse thermal energy $\beta=1/(k_{\rm B}T)$ and the electron charge $e$.}

\subsection{Test-charge theory}

\RP{Here we briefly} review the \RP{\sl test charge approach} developed in Ref.~\cite{PRE2019}. In order to characterize the thermodynamic equilibrium state of the polymer-membrane complex,  we will use the field-theoretic formulation of \RP{the partition function of a} charged systems. Within this formalism, the grand-canonical partition function of the electrolyte is given by a functional integral over the fluctuating electrostatic potential $\phi(\br)$~\cite{Podgornik88}, 
\be\label{zg2}
Z_{\rm G}=\int \mathcal{D}\phi\;e^{-H[\phi]}, 
\ee
where the \SB{dimensionless} Hamiltonian functional reads
\bea\label{HamFunc}
H[\phi]&=&\frac{k_{\rm B}T}{2e^2}\int\mathrm{d}\br\;\e(\br)\left[\nabla\phi(\br)\right]^2-i\int\mathrm{d}\br\sigma(\br)\phi(\br)\nonumber\\
&&-\sum_{i=\{\pm,c\}}\Lambda_i \int\mathrm{d}\br\;e^{iq_i\phi(\br)}\theta_{\rm s}(z).
\eea
In Eq.~(\ref{HamFunc}), the first term on the r.h.s. corresponds to the free energy of the pure solvent. The second term incorporates the total macromolecular charge density
\be\label{mc}
\sigma(\br)=\sigma_{\rm m}(\br)+\sigma_\p(\br),
\ee
where $\sigma_\p(\br)$ stands for the polymer charge density function. Finally, the third term of Eq.~(\ref{HamFunc}) is the fluctuating mobile ion density, with the index symbols $i=\{+,-,c\}$ corresponding respectively to the monovalent cations  and anions ($q_\pm=\pm1$), and multivalent counterions. From now on, monovalent ions will be called \textit{salt} while multivalent cations will be simply designated by \textit{counterions}. We also note that the Heaviside step function $\theta_{\rm s}(z)$ in Eq.~(\ref{HamFunc}) restricts the ion partition to the upper half space $z>0$ bounded by the impenetrable membrane. 

The test charge approach is based on the perturbative treatment of the polyelectrolyte with the \RP{explicit aim} to {\sl recover the planar symmetry} broken by the polymer molecule. Within this approximation, we Taylor expand the partition function~(\ref{zg2}) up to the quadratic order in the polymer charge density $\sigma_\p(\br)$ to get
\bea\label{zg3}
Z_{\rm G}&=&Z_0\left\{1+i\int\mathrm{d}\br\sigma_{\RP{\p}}(\br)\lan\phi(\br)\ran_0\right.\\
&&\left.\hspace{7mm}-\frac{1}{2}\int\mathrm{d}\br\mathrm{d}\br'\sigma_\p(\br)\lan\phi(\br)\phi(\br')\ran_0\sigma_\p(\br')\right\},\nonumber
\eea
with the polymer-free partition function 
\be\label{z0}
Z_0=\int \mathcal{D}\phi\;e^{-H_0[\phi]}
\ee
including the Hamiltonian
\bea\label{H0}
H_0[\phi]&=&\frac{k_{\rm B}T}{2e^2}\int\mathrm{d}\br\;\e(\br)\left[\nabla\phi(\br)\right]^2-i\int\mathrm{d}\br\sigma_{\rm m}(\br)\phi(\br)\nonumber\\
&&-\sum_{i=\{\pm,c\}}\Lambda_i \int\mathrm{d}\br\;e^{iq_i\phi(\br)}\theta_{\rm s}(z).
\eea
The brackets in Eq.~(\ref{zg3}) denote the field average with the polymer-free Hamiltonian~(\ref{H0}), i.e.
\be
\label{br}
\lan F[\phi]\ran_0=\frac{1}{Z_0}\int \mathcal{D}\phi\;e^{-H_0[\phi]}F[\phi].
\ee

The computation of the electrostatic grand potential $\beta\Omega_{\rm G}\equiv-\ln Z_{\rm G}$ at the same quadratic order in the polymer charge $\sigma_\p(\br)$ yields
\bea\label{og}
\beta\Omega_{\rm G}&=&\beta\Omega_0+\int\mathrm{d}\br\sigma_\p(\br)\bph(\br)\\
&&+\frac{1}{2}\int\mathrm{d}\br\mathrm{d}\br'\sigma_\p(\br)G(\br,\br')\sigma_\p(\br'),\nonumber
\eea
with the grand potential of the polymer-free electrolyte $\beta\Omega_0=-\ln Z_0$, and the average value of the electrostatic potential $\phi(\br)$ and its two-point correlation function,
\bea
\label{pm}
\bph(\br)&=&-i\lan\phi(\br)\ran_0,\\
\label{gr}
G(\br,\br')&=&\lan\phi(\br)\phi(\br')\ran_0-\lan\phi(\br)\ran_0\lan\phi(\br')\ran_0.
\eea
From Eq.~(\ref{og}), the polymer grand potential $\Omega_\p=\Omega_{\rm G}-\Omega_0$ corresponding to the net contribution from the polyelectrolyte charge to the total grand potential  follows as
\be\label{ogp}
\beta\Omega_\p=\int\mathrm{d}\br\sigma_\p(\br)\bph(\br)+\frac{1}{2}\int\mathrm{d}\br\mathrm{d}\br'\sigma_\p(\br)G(\br,\br')\sigma_\p(\br').
\ee

\subsection{Evaluating the polymer grand potential within the \RP{$1\ell$}-dressed SC theory}
\label{sccoun}

Due to the non-linearity of the Hamiltonian~Eq. (\ref{H0}), the field averages in Eqs.~(\ref{pm}) and~(\ref{gr}) cannot be evaluated exactly. Thus, we introduce here a \RP{$1\ell$}-corrected SC formalism that will enable us to evaluate analytically the grand potential~Eq. (\ref{ogp}). To this end, we first recast the Hamiltonian functional~(\ref{H0}) in the form
\be\label{H02}
H_0[\phi]=H_{\rm s}[\phi]+H_{\rm c}[\phi], 
\ee
with the Hamiltonian of the monovalent salt and the multivalent counterions 
\bea
\label{Hs}
H_{\rm s}[\phi]&=&\frac{k_{\rm B}T}{2e^2}\int\mathrm{d}\br\;\e(\br)\left[\nabla\phi(\br)\right]^2-i\int\mathrm{d}\br\sigma_{\rm m}(\br)\phi(\br)\nonumber\\
&&-\int\mathrm{d}\br\left[\Lambda_+e^{i\phi(\br)}+\Lambda_-e^{-i\phi(\br)}\right]\theta_{\rm s}(z),\\
\label{Hq}
H_{\rm c}[\phi]&=&-\Lambda_\C\int\mathrm{d}\br_\C\;e^{iq_{\rm c}\phi(\br_\C)}\theta_{\rm s}(z_\C).
\eea
From now on, the counterion coordinates will be denoted by the position vector $\br_\C=(x_\C,y_\C,z_\C)$. Due to their high valency resulting in strong correlations with the membrane charges, these counterions will be treated within the SC approximation equivalent to a fugacity expansion~\cite{NetzSC}. This low fugacity approximation is also motivated by the fact that in experiments, the bulk tri- and tetravalent counterion concentration is by orders of magnitude lower than the bulk monovalent salt concentration~\cite{Podgornik2010}. Thus, Taylor-expanding the functional integral and the partition function $Z_0$ of Eq.~(\ref{br}) in terms of the fugacity $\Lambda_{\rm c}$, one gets the SC-expanded field average of the general functional $F[\phi]$ in the form
\bea
\label{avex}
\lan F[\phi]\ran_0&=&\lan F[\phi]\ran_{\rm s}+\Lambda_{\rm c}\int\mathrm{d}\br_\C\left\{\lan F[\phi]e^{iq_{\rm c}\phi(\br_\C)}\ran_{\rm s}\right.\\
&&\hspace{2.cm}\left.-\lan F[\phi]\ran_{\rm s}\lan e^{iq_{\rm c}\phi(\br_\C)}\ran_{\rm s}\right\}\theta_{\rm s}(z_\C),\nonumber
\eea
where we defined the field average with the Hamiltonian of the salt ions in Eq.~(\ref{Hs}),
\be
\label{brs}
\lan F[\phi]\ran_\s=\frac{1}{Z_s}\int \mathcal{D}\phi\;e^{-H_s[\phi]}F[\phi],
\ee
with the salt partition function $Z_\s=\int \mathcal{D}\phi\;e^{-H_\s[\phi]}$. 

The interactions  of the monovalent ions with the membrane charges are characterized by weak to intermediate electrostatic coupling~\cite{Buyuk2012}. Thus, in the following, the salt characterized by the Hamiltonian~(\ref{Hs}) will be treated at \RP{$1\ell$}-level. In other words, Eq.~(\ref{Hs}) will be approximated by an Hamiltonian quadratic in the fluctuating potential $\phi(\br)$, with the average value and variance corresponding respectively to the \RP{$1\ell$}-level mean electrostatic potential $\phi_{\rm m}(\br)$ and correlation function $v(\br,\br')$,
\be\label{gaus}
H_{\rm s}[\phi]\approx\frac{1}{2}\int_{\br,\br'}\left[\phi(\br)-i\phi_{\rm m}(\br)\right]v(\br,\br')\left[\phi(\br')-i\phi_{\rm m}(\br')\right]. 
\ee
Eqs.~(\ref{brs}) and~(\ref{gaus}) yield indeed the expectation values
\bea\label{eqs3}
\lan\phi(\br)\ran_\s&=&i\phi_{\rm m}(\br),\\
\label{eqs4}
\lan\phi(\br)\phi(\br')\ran_\s&=&v(\br,\br')-\phi_{\rm m}(\br)\phi_{\rm m}(\br').
\eea

Evaluating now the average potential~(\ref{pm}) and the correlator~(\ref{gr}) with Eqs.~(\ref{avex})-(\ref{gaus}), after some algebra, one obtains
\bea
\label{eqs4I}
&&\bph(\br)=\phi_{\rm m}(\br)+q_\C\int\mathrm{d}\br_\C v(\br,\br_\C)\rho_{\rm c}(\br_\C),\\
\label{eqs4II}
&&G(\br,\br')=v(\br,\br')-q_\C^2\int\mathrm{d}\br_\C v(\br,\br_\C)\rho_{\rm c}(\br_\C)v(\br_\C,\br'),\nonumber\\
\eea
where we introduced the counterion density
\be\label{eqs10}
\rho_{\rm c}(\br_\C)=\Lambda_{\rm c}\;e^{-\frac{q_{\rm c}^2}{2}v(\br_\C,\br_\C)-q_{\rm c}\phi_{\rm m}(\br_\C)}\theta_{\rm s}(z_\C).
\ee
The corresponding \textit{\RP{$1\ell$}-dressed SC theory} is a generalized SC approach which assumes that the interactions of multivalent counterions with the membrane charges are subjected to the non-uniform screening by the monovalent salt whose spatial density variation is taken into account at the non-linear \RP{$1\ell$}-level. Thus, the present approach upgrades the \textit{dressed ion theory} of Refs.~\cite{Podgornik2010,Podgornik2011} that treats the salt ions at the linear DH-level. 

Finally, substituting Eqs.~(\ref{eqs4I}) and~(\ref{eqs4II}) into Eq.~(\ref{ogp}), the polymer grand potential follows as
\be
\label{eqs5}
\Omega_{\rm p}=\Omega_{\rm pm}+\Omega_{\rm pp}+\Omega_{\rm p{\rm c}}+\Omega_{\rm p{\rm c}p},
\ee
with the interaction potential components
\bea
\label{eqs6}
\beta\Omega_{\rm pm}&=&\int\mathrm{d}\br\sigma_{\rm p}(\br)\phi_{\rm m}(\br),\\
\label{eqs7}
\beta\Omega_{\rm pp}&=&\frac{1}{2}\int\mathrm{d}\br\mathrm{d}\br'\sigma_{\rm p}(\br)v(\br,\br')\sigma_{\rm p}(\br'),\\
\label{eqs8}
\beta\Omega_{\rm pc}&=&q_{\rm c}\int\mathrm{d}\br\mathrm{d}\br_\C\sigma_{\rm p}(\br)v(\br,\br_\C)\rho_{\rm c}(\br_\C),\\
\label{eqs9}
\beta\Omega_{\rm pcp}&=&-\frac{q_{\rm c}^2}{2}\int_{\br,\br',\br_\C}\sigma_{\rm p}(\br)v(\br,\br_\C)\rho_{\rm c}(\br_\C)v(\br_\C,\br')\sigma_{\rm p}(\br').\nonumber\\
\eea
The physical meaning of the energy components~(\ref{eqs6})-(\ref{eqs9}) have been qualitatively emphasized below Eq.~(\ref{EqInt}). Eqs.~(\ref{eqs6}) and Eqs.~(\ref{eqs7}) correspond respectively to the direct polymer-membrane charge coupling energy, and the polymer self-energy originating from the non-uniform screening of the polymer charges by the spatially varying salt strength. These two potential components have been previously derived in Ref.~\cite{PRE2019}. Eq.~(\ref{eqs8}) is in turn the direct polymer-counterion interaction energy. Finally, Eq.~(\ref{eqs9}) is a salt-dressed three-body potential that accounts for the screening of the polymer self-energy by the inhomogeneously distributed multivalent counterions.

\subsection{The planar symmetry}
\label{ps}

In order to simplify the coupling potentials in Eqs.~(\ref{eqs6})-(\ref{eqs9}), we now account for the planar symmetry of the membrane characterized by the equalities 
\bea
\label{potsym}
\phi_{\rm m}(\br)&=&\phi_{\rm m}(z)\\
\label{grsym}
v(\br,\br')&=&\int\frac{d^2\bk}{4\pi^2}e^{i\bk\cdot\left(\br_{\pa}-\br'_{\pa}\right)}\tv(z,z';k).
\eea
In the Fourier transform of the Green's function in Eq.~(\ref{grsym}), we used the translational symmetry of the electrostatic interactions along the membrane surface, i.e. $v(\br,\br')=v(\br_{\pa}-\br'_{\pa},z,z')$, with the position vector $\br_{\pa}=x\hat{u}_x+y\hat{u}_y$ in the $x-y$ plane. Using Eqs.~(\ref{potsym})-(\ref{grsym}) in Eq.~(\ref{eqs10}),  the counterion density simplifies as
\be
\label{densym}
\rho_{\rm c}(\br_\C)=\rho_{\rm c}(z_\C)=\Lambda_{\rm c}\;e^{-\frac{q_{\rm c}^2}{2}v\left(\br_{\pa}-\br'_{\pa}=\mathbf{0},z_\C,z_\C\right)-q_{\rm c}\phi_{\rm m}(z_\C)}\theta_{\rm s}(z_\C).
\ee
In order to determine the counterion fugacity $\Lambda_{\rm c}$, we evaluate Eq.~(\ref{densym}) in the bulk region $z\to\infty$ where $\phi_{\rm m}(z)\to0$ and $v(\br,\br')\to v_{\rm b}(\br-\br')$, with the \RP{$1\ell$}-level bulk Green's function given by the screened Coulomb potential~\cite{Buyuk2012}
\be
\label{18}
v_{\rm b}(\br-\br')=\ell_{\rm B}\frac{e^{-\kappa|\br-\br'|}}{|\br-\br'|}.
\ee
This yields $\Lambda_{\rm c}=\rho_{{\rm bc}}\;e^{\frac{q_{\rm c}^2}{2}v_{\rm b}(\br-\br')}|_{\br'\to\br}$, and the counterion density~(\ref{eqs10}) finally follows as
\be\label{eqs11}
\rho_{\rm c}(z_\C)=\rho_{{\rm bc}}\;e^{-\frac{q_{\rm c}^2}{2}\delta v(z_\C)-q_{\rm c}\phi_{\rm m}(z_\C)}\theta_{\rm s}(z_\C),
\ee
where defined the ionic self-energy corresponding to the renormalized equal point correlation function
\be\label{ionself}
\delta v(z)\equiv \lim_{\br'\to\br} \left\{v\left(\br_{\pa}-\br'_{\pa},z,z\right)-v_{\rm b}(\br-\br')\right\}.
\ee
In Eq.~(\ref{18}), we used the Bjerrum length $\ell_{\rm B}= e^2/(4\pi\SB{\e_0}\e_{\rm w} k_{\rm B}T)\approx7$ {\AA} corresponding to the separation distance where two point ions interact with the thermal energy $k_{\rm B}T$, and the DH screening parameter $\kappa=\sqrt{8\pi\ell_{\rm B}\rho_{\rm b}}$ whose inverse gives the characteristic radius of the monovalent counterion cloud around a bulk ion. The electrostatic model parameters are summarized in Table I. Using now Eqs.~(\ref{potsym})-(\ref{eqs11}), the interaction potentials in Eqs.~(\ref{eqs6})-(\ref{eqs9}) simplify to
\bea\label{eqs12}
\beta\Omega_{\rm pm}(z_\p,\te)&=&-\tau\int_{-L/2}^{L/2}\md l\;\phi_{\rm m}\left(z_\p+l\ct\right),\\
\label{eqs13}
\beta\Omega_{\rm pp}(z_\p,\te)&=&\frac{\tau^2}{2}\int\frac{\md\bk}{4\pi^2}\int_{-L/2}^{L/2}\md l\int_{-L/2}^{L/2}\md l'e^{i\bk\cdot(\bl-\bl')}\nonumber\\
&&\hspace{.4cm}\times\tv\left(z_\p+l\ct,z_\p+l'\ct;k\right),\nonumber\\
&&\\
\label{eqs14}
\beta\Omega_{\rm pc}(z_\p,\te)&=&-q_{\rm c}\tau \int_{-L/2}^{L/2}\md l\int_0^\infty\md z_\C\rho_{\rm c}(z_\C)\\
&&\hspace{1cm}\times\tv(z_{\rm p}+l\ct,z_\C;k=0),\nonumber\\
\label{eqs15}
\beta\Omega_{\rm pcp}(z_\p,\te)&=&-\frac{(q_{\rm c}\tau)^2}{2}\int\frac{\md\bk}{4\pi^2}\int_0^\infty\md z_\C\rho_{\rm c}(z_\C)\\
&&\times\left|\int_{-L/2}^{L/2}\md l\;e^{i\bk\cdot\bl}\;\tv(z_{\rm p}+l\ct,z_\C;k)\right|^2,\nonumber
\eea
where we defined the scalar product $\bk\cdot\bl=kl\st\cpk$.

The net electrostatic energy characterizing the nature of the polymer-membrane interactions corresponds to the polymer grand potential~(\ref{eqs5}) renormalized by its bulk limit, 
\be\label{netpot}
\Delta\Omega_{\rm p}(z_\p,\te)=\Omega_{\rm p}(z_\p,\te)-\Omega_{\rm p,b},
\ee
with the bulk grand potential  
\be\label{bulkpot}
\Omega_{\rm p,b}=\lim_{z_{\p}\to\infty}\Omega_{\rm p}(z_\p,\te).
\ee
Eq.~(\ref{bulkpot}) corresponds to the adiabatic work to be done on the polymer in order to bring the molecule from the bulk reservoir to the distance $z_{\rm p}$ from the membrane. In terms of the grand potential~(\ref{netpot}), the orientation-averaged polymer number density is given by
\be\label{dn}
\rho_{\rm p}(z)=\frac{\rho_{\rm pb}}{2}\int_{\theta_-}^{\theta_+}\mathrm{d}\theta\sin\theta e^{-\beta\Delta\Omega_\p(z_\p,\te)}
\ee
where we introduced the bulk polymer concentration $\rho_{\rm bp}$. Finally, the average polymer orientation can be characterized by the \RP{\sl orientational order parameter}
\be\label{or}
S_\p(z_\p)=\frac{3}{2}\left[\lan\cos^2\theta_\p\ran-\frac{1}{3}\right],
\ee
with the orientational average  defined as
\be\label{defi}
\lan f(\theta_\p)\ran=\frac{\int_{\theta_-}^{\theta_+}\mathrm{d}\theta_\p\sin\theta_\p f(\theta_\p) e^{-\beta\Delta\Omega_\p(z_\p,\te)}}{\int_{\theta_-}^{\theta_+}\mathrm{d}\theta_\p\sin\theta_\p e^{-\beta\Delta\Omega_\p(z_\p,\te)}}.
\ee
The value $S_\p(z_{\rm p})=-1/2$ corresponds to the parallel polymer orientation with the membrane and $S_\p(z_{\rm p})=1$ indicates the perpendicular configuration of the molecule. For vanishing electrostatic interactions  $\Delta\Omega_\p(z_\p,\te)=0$, and in the absence of steric penalty where $\theta_-=0$ and $\theta_+=\pi$, the \RP{\sl orientational order parameter} yields $S_\p(z_{\rm p})=0$ indicating the freely rotating polymer regime. Finally, in the presence of steric penalty without electrostatic interactions,  the polymer density~(\ref{dn}) and \RP{\sl orientational order parameter}~(\ref{or}) take the piecewise form
\bea\label{dest}
\rho_{\rm p}(z_\p)&=&\rho_{\rm pb}\hspace{1mm}\mathrm{min}\left(1,\frac{2z_\p}{L}\right),\\
\label{orst}
S_\p(z_\p)&=&\frac{1}{2}\hspace{1mm}\mathrm{min}\left(0,\frac{4z_\p^2}{L^2}-1\right).
\eea
Thus, over the interfacial region $0\leq z_\p\leq L/2$, the polymer density grows linearly and the \RP{\sl orientational order parameter} quadratically towards their bulk value $\rho_\p(z_\p)=\rho_{\rm b}$ and $S_\p(z_{\rm p})=0$. Eq.~(\ref{orst}) is plotted in the inset of Fig.~\ref{fig2}(b) (see the thin solid curve).

\begin{table}[ht]
\caption{Model Parameters and Coupling Constants}
\begin{tabular}{c c}
\hline\hline 
Monovalent salt concentration & $\rho_{\rm b}$\\ [1.0 ex] 
\hline
Multivalent cation concentration & $\rho_{\rm bc}$\\ [1.0 ex] 
\hline
Multivalent cation valency & $q_\C$\\ [1.0 ex] 
\hline
Membrane charge density & $-\sigma_{\rm m}$\\ [1.0 ex] 
\hline
Polymer charge density & $-\tau$\\ [1.0 ex] 
\hline
Bjerrum length & $\ell_{\rm B}=\frac{e^2}{4\pi\SB{\e_0}\ew k_{\rm B}T}\approx 7$ {\AA}\\ [1.0 ex] 
\hline
GC length & $\mu=1/(2\pi\ell_{\rm B}\sigma_{\rm m})$\\ [1.0 ex] 
\hline
Salt screening parameter & $\kappa=\sqrt{8\pi\ell_{\rm B}\rho_{\rm b}}$\\ [1.0 ex] 
\hline
Relative screening strength & $s=\kappa\mu$\\ [1.0 ex] 
\hline
Auxiliary screening parameter & $\gamma=\sqrt{s^2+1}-s$\\ [1.0 ex] 
\hline
Monovalent cation coupling strength & $\Xi_{\rm c}=\frac{\ell_{\rm B}}{\mu}$\\ [1.0 ex]
\hline
Monovalent salt coupling strength & $\Gamma_{\rm s}=\kappa\ell_{\rm B}=s\Xi_{\rm c}$\\ [1.0 ex]
\hline
Multivalent cation coupling strength & $\Gamma_\C=\frac{q_{\rm c}^2\rho_{\rm bc}}{4\rho_{\rm b}}$.\\ [1.0 ex]
\hline
\end{tabular}
\label{table:nonlin}
\end{table}

\section{Results}
\label{res}

In this Section, polymer-membrane interactions will be characterized in terms of the electrostatic polymer grand potential~(\ref{netpot}). The evaluation of this grand potential requires the determination of the monovalent salt-dressed average electrostatic potential $\phi_{\rm m}(z)$ and correlator $v(\br,\br')$ in Eqs.~(\ref{eqs11})-(\ref{eqs15}). As explained in Section~\ref{sccoun}, this task will be achieved within the \RP{$1\ell$} theory of electrostatic interactions. According to the \RP{$1\ell$} theory of asymmetrically partitioned salt solutions, the average potential and the Fourier-transformed correlator read~\cite{Buyuk2012}
\bea\label{potsup1}
\phi_{\rm m}(z)&=&\phi^{(0)}_{\rm m}(z)+\phi^{(1)}_{\rm m}(z),\\
\label{potsup2}
\tv(z,z';k)&=&\tv_{\rm b}(z-z';k)+\delta\tv\left(z,z';k\right).
\eea
The first terms on the r.h.s. of Eqs.~(\ref{potsup1}) and~(\ref{potsup2}) correspond respectively to the solution of the MF-level PB equation 
\be\label{PB}
\partial_z^2\phi^{(0)}_{\rm m}(z)-\kappa^2\sinh\left[\phi^{(0)}_{\rm m}(z)\right]=4\pi\ell_{\rm B}\sigma_{\rm m}\delta(z),
\ee
and the Fourier transform of the bulk Green's function~(\ref{18}), 
\be
\label{tvb}
\tv_{\rm b}(z-z';k)=\frac{2\pi\ell_{\rm B}}{p}e^{-p|z-z'|},
\ee
with the screening function $p=\sqrt{k^2+\kappa^2}$. The PB Eq.~(\ref{PB}) is solved by the potential function~\cite{Isr}
\be\label{10}
\phi^{(0)}_{\rm m}(z)=-2\ln\left[\frac{1+\gamma e^{-\kappa z}}{1-\gamma e^{-\kappa z}}\right],
\ee
where we used the auxiliary coefficient  $\gamma=\sqrt{s^2+1}-s$. The parameter $s=\kappa\mu$ involves the Gouy-Chapman (GC) length $\mu=1/(2\pi\ell_{\rm B}\sigma_{\rm m})$ corresponding to the characteristic thickness of the interfacial monovalent cations. Thus, this parameter scaling as $s\propto\sigma_{\rm m}^{-1}\rho^{1/2}_{\rm b}$ measures the relative density and screening strength of the interfacial monovalent cations and the bulk salt.

The second potential terms on the r.h.s. of Eqs.~(\ref{potsup1}) and~(\ref{potsup2}) bring membrane-salt correlations of \RP{$1\ell$} order~\cite{Buyuk2012}. The computation of these correlation potentials derived in Ref.~\cite{Buyuk2012} is explained in Appendix~\ref{pots1l}. Therein, we show that the corresponding potentials  scale as $\phi^{(1)}_{\rm m}\propto\Gamma_{\rm s}$ and $\delta v\propto\Gamma_{\rm s}$, with the electrostatic coupling parameter $\Gamma_{\rm s}=\kappa\ell_{\rm B}$ measuring the importance of salt fluctuations. This parameter is related to the interfacial monovalent counterion coupling parameter $\Xi_{\rm c}=\ell_B/\mu$ of Ref.~\cite{NetzSC} as $\Gamma_{\rm s}=s\Xi_{\rm c}$ (see Table I). 

Section~\ref{mfsalt} deals with polymer-membrane interactions in the regime of weak membrane charges where these correlation corrections are perturbative. Thus, in Section~\ref{mfsalt}, the salt distribution is treated at MF level. In Section~\ref{1lsalt}, this analysis is extended to the case of intermediate membrane charges where the emerging salt-membrane correlations are taken into account within the electrostatic \RP{$1\ell$} theory. The \RP{$1\ell$}-level evaluation of the grand potential components in Eqs.~(\ref{eqs12})-(\ref{eqs15}) is explained in Appendix~\ref{1lpol}.

\subsection{Like-charge complexation of weakly charged polymers and membranes: MF salt}
\label{mfsalt}

We investigate here the alteration of the MF-level like-charge polymer-membrane repulsion by the exclusive effect of the multivalent counterions. To this end, we focus on the regime of weak monovalent salt $\Gamma_{\rm s}<1$ and low membrane charges $s>1$ where the monovalent ion-membrane correlations measured by the coupling parameter $\Xi_{\rm c}=\Gamma_{\rm s}/s<1$ are negligible.  Moreover, we consider a weak polymer charge and set $\bt=0.05$. Thus, we treat the salt distribution at the MF-level, and also neglect the polymer self-energy potentials~(\ref{eqs13}) and~(\ref{eqs15}) carrying salt-induced correlations and second order (quadratic) polymer charge corrections. Within this MF approximation, the polymer grand potential~(\ref{netpot}) simplifies to
\be\label{eqs16}
\Delta\Omega^{(0)}_{\rm p}(z_\p,\te)=\Omega^{(0)}_{\rm pm}(z_\p,\te)+\Delta\Omega^{(0)}_{\rm pc}(z_\p,\te).
\ee

The first term of Eq.~(\ref{eqs16}) is the MF-level polymer-membrane coupling energy. Substituting the MF potential~(\ref{10}) into Eq.~(\ref{eqs12}), this interaction energy follows as
\bea
\label{11}
\beta\Omega^{(0)}_{\rm pm}(\tz_\p,\te)&=&\frac{2\tau}{\kappa\ct}\left\{\mathrm{Li}_2\left[\gamma e^{-\tz_-}\right]-\mathrm{Li}_2\left[-\gamma e^{-\tz_-}\right]\right.\\
&&\hspace{1.4cm}\left.-\mathrm{Li}_2\left[\gamma e^{-\tz_+}\right]+\mathrm{Li}_2\left[-\gamma e^{-\tz_+}\right]\right\}\nonumber
\eea
where we used the polylog function $\mathrm{Li}_2(x)$~\cite{math}, and the distance between the polymer edges and the membrane,
\be\label{11II}
\tz_\pm=\tz_\p\pm\frac{\tL}{2}\ct,
\ee
with the dimensionless CM distance $\tz_\p=\kappa z_\p$ and polymer length $\tL=\kappa L$. In the strict MF DH regime of weak membrane charges where $s\gg1$, Eq.~(\ref{11II}) simplifies to
\be\label{12}
\beta\Omega^{(0)}_{\rm pm}(\tz_\p,\te)\approx\frac{2}{s} \tau L_\p(\te)e^{-\tz_\p}
\ee
with the effective polymer length 
\be
\label{13}
L_\p(\te)=\frac{2\sinh\left(\tL\ct/2\right)}{\kappa\ct}.
\ee 
Eq.~(\ref{12}) indicates that MF-level polymer-membrane coupling is characterized by purely repulsive interactions. Due to screening by salt, these interactions decay  exponentially with the polymer distance.

The second term of Eq.~(\ref{eqs16}) corresponds to the normalized polymer-counterion interaction potential 
\be\label{eqs16II}
\Delta\Omega^{(0)}_{\rm pc}(z_\p,\te)=\Omega^{(0)}_{\rm pc}(z_\p,\te)-\Omega_{\rm pc,b}, 
\ee
with the MF limit of Eq.~(\ref{eqs14})
\bea\label{eqs17}
\beta\Omega^{(0)}_{\rm pc}(z_\p,\te)&=&-q_{\rm c}\tau \int_{-L/2}^{L/2}\md l\int_0^\infty\md z_\C\rho^{(0)}_{\rm c}(z_\C)\\
&&\hspace{1cm}\times\tv_{\rm b}(z_{\rm p}+l\ct,z_\C;k=0),\nonumber
\eea
and its bulk value computed in Appendix~\ref{1lpc},
\be\label{pcb}
\beta\Omega_{\rm pc,b}=-\frac{4\pi\ell_{\rm B}\rho_{\rm bc}}{\kappa^2}L\tau\qc=-\frac{2\Gamma_{\rm c}}{\qc}L\tau.
\ee
In Eq.~(\ref{pcb}), we introduced the additional coupling parameter $\Gamma_{\rm c}$ characterizing the competition between the counterions and salt,
\be\label{eqs22}
\Gamma_{\rm c}\equiv\frac{2\pi\qc^2\ell_{\rm B}\rho_{\rm bc}}{\kappa^2}=\frac{q_{\rm c}^2\rho_{\rm bc}}{4\rho_{\rm b}}.
\ee
We emphasize that in Eq.~(\ref{eqs17}), correlations associated with salt were neglected by replacing the Green's function $\tv(z,z';k)$ in Eqs.~(\ref{eqs14}) by its bulk component~(\ref{tvb}). Moreover, we included the counterion density~(\ref{eqs11}) evaluated at the MF-level,
\be\label{eqs18}
\rho^{(0)}_{\rm c}(z_\C)=\rho_{{\rm bc}}e^{-q_{\rm c}\phi^{(0)}_{\rm m}(z_\C)}=\rho_{{\rm bc}}\left(\frac{1+\gamma e^{-\kappa z_\C}}{1-\gamma e^{-\kappa z_\C}}\right)^{2q_{\rm c}},
\ee
where we used the MF average potential~(\ref{10}). Finally, carrying-out the double integral in Eq.~(\ref{eqs17}), one gets
\be\label{eqs19}
\beta\Delta\Omega^{(0)}_{\rm pc}(z_\p,\te)=-\frac{2\pi\ell_{\rm B}\tau\rho_{\rm bc}\qc}{\kappa^3\ct}\Psi(z_\p,\te),
\ee
with the auxiliary function
\bea
\label{eqs19II}
\Psi(z_\p,\te)&=&e^{-\zm}\left[J_1(\zm)-J_1(0)\right]-e^{-\zp}\left[J_1(\zp)-J_1(0)\right]\nonumber\\
&&+e^{\zm}\left[J_{-1}(\zm)-J_{-1}(\infty)\right]\nonumber\\
&&-e^{\zp}\left[J_{-1}(\zp)-J_{-1}(\infty)\right]\nonumber\\
&&+2\left[J_0(\zp)-J_0(\zm)-\tL\ct\right].
\eea
In Eq.~(\ref{eqs19II}), we used the dimensionless coordinates defined in Eq.~(\ref{11II}) and introduced the integral function
\be\label{eqs20}
J_n(x)=\int\mathrm{d}x\;e^{nx}\left(\frac{1+\gamma e^{-x}}{1-\gamma e^{-x}}\right)^{2q_{\rm c}}
\ee
whose explicit form is given in Appendix~\ref{apI}.

\begin{figure}
\includegraphics[width=1\linewidth]{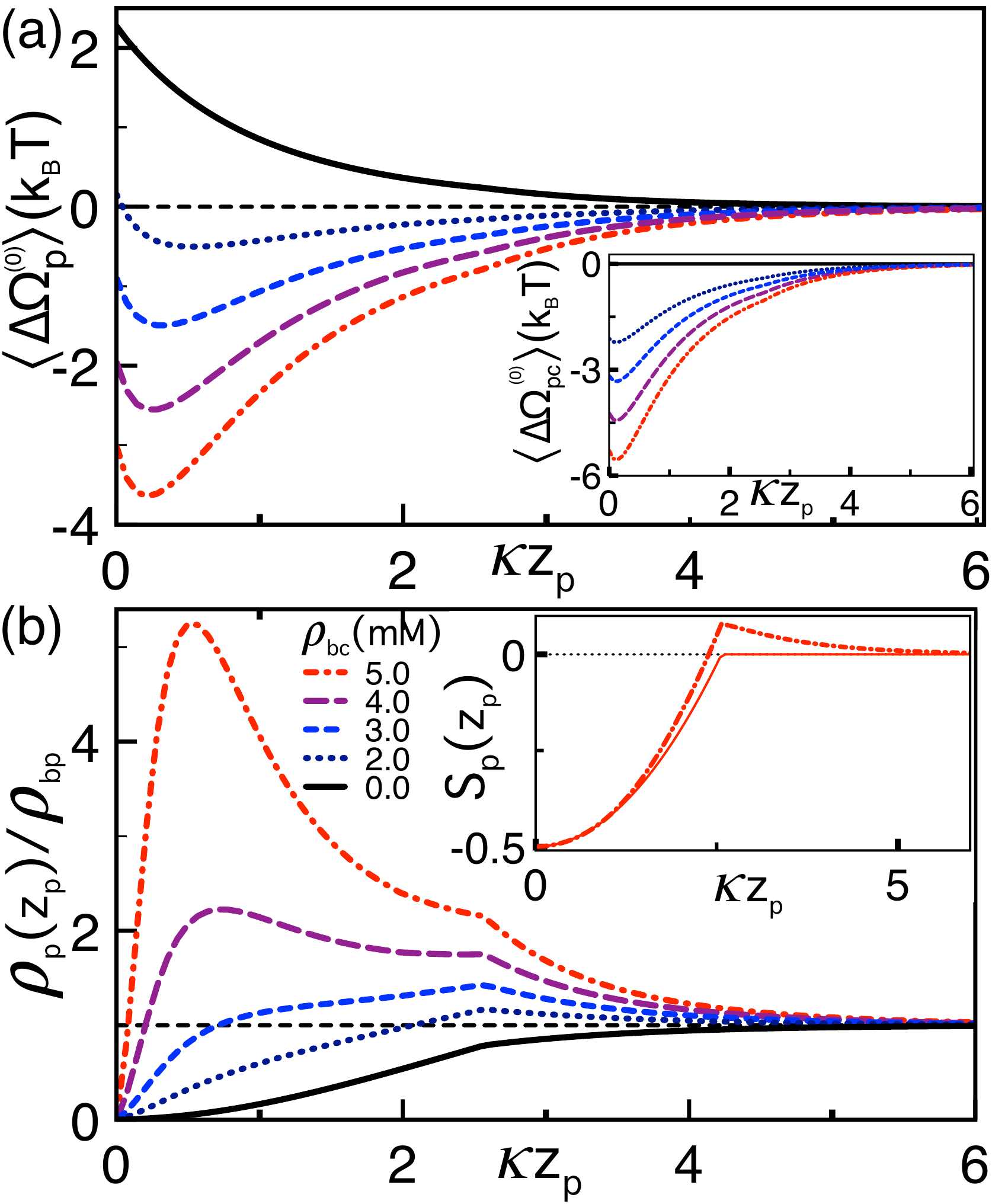}
\caption{(Color online)  (a) Grand potential~(\ref{eqs16}) (main plot) and polymer-counterion coupling potential~(\ref{eqs19}) (inset) averaged over polymer rotations. (b) Polymer density~(\ref{dn}) (main plot) and \RP{\sl orientational order parameter}~(\ref{or}) (inset) computed with the MF grand potential~(\ref{eqs16}). The thin solid curve in the inset is from Eq.~(\ref{orst}). The bulk density of the tetravalent counterions ($q_{\rm c}=4$) is given in the legend of (b). The membrane charge is $\sigma_{\rm m}=0.2$ $e/\rm{nm}^2$, polymer length $L=5$ nm and charge $\bt=0.05$, and salt concentration $\rho_{\rm b}=0.1$ M.}
\label{fig2}
\end{figure}
\begin{figure*}
\includegraphics[width=1\linewidth]{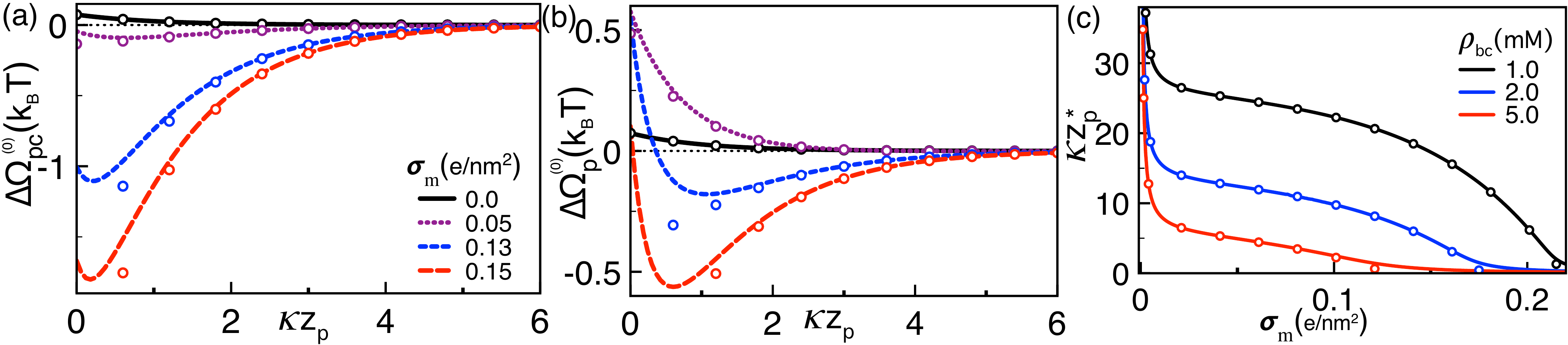}
\caption{(Color online) Curves: (a)  Polymer-tetravalent counterion interaction potential~(\ref{eqs19}) and (b) total grand potential~(\ref{eqs16}) at the tetravalent counterion concentration $\rho_{\rm bc}=5.0$ mM. The circles are the asymptotic limits in Eqs.~(\ref{eqs21}) and~(\ref{eqs23}). (c) The binding position of the polymer obtained from the minimum of the grand potential~(\ref{eqs16}) (curves) and Eq.~(\ref{eqs25}) (circles). The salt concentration is $\rho_{\rm b}=0.1$ M and  the polymer angle $\te=\pi/2$. The other parameters are the same as in Fig.~\ref{fig2}.}
\label{fig3}
\end{figure*}

\subsubsection{Onset of like-charge polymer adsorption by multivalent counterion addition}

We characterize here the experimental observation of like-charge polymer adsorption by multivalent cation addition~\cite{Caball2014,Tiraferri2015,Fries2017}. Figs.~\ref{fig2}(a) and (b) display the polymer grand potential and density at various tetravalent counterion concentrations. One sees that in the absence of counterions (black curves), like-charge polymer-membrane interactions lead to a purely repulsive polymer grand potential ($\Delta\Omega^{(0)}_{\rm p}>0$) and a total polymer depletion from the interface ($\rho_{\rm p}<\rho_{\rm bp}$). Then, the inset of Figs.~\ref{fig2}(a) shows that the presence of tetravalent counterions leads to an attractive polymer-counterion interaction potential $\Delta\Omega^{(0)}_{\rm pc}<0$. One notes that this effect is amplified by further counterion addition, i.e. $\rho_{\rm bc}\uparrow\Delta\Omega^{(0)}_{\rm pc}\downarrow$. Consequently, close to the membrane, the total polymer grand potential $\Delta\Omega^{(0)}_{\rm p}$ develops an attractive well, indicating the onset of like-charge polymer attraction by the membrane surface. This results in a polymer adsorption peak that rises with the counterion density, i.e. $\rho_{\rm bc}\uparrow\Delta\Omega^{(0)}_{\rm p}\downarrow\rho_{\rm p}\uparrow$. 

The inset of Fig.~\ref{fig2}(b) shows that in the interfacial region $\tz_{\rm p}<\tL/2\approx2.5$ governed by the steric penalty, the \RP{\sl orientational order parameter} (dashed curve) remains close to the pure steric limit characterized by the parallel polymer orientation $S_{\rm p}(\tz_{\rm p})<0$ (thin solid curve). Thus, close to the membrane surface, the like-charge attraction and the subsequent adsorption of the polymer occurs in the parallel configuration of the molecule. However, one notes that in the outer region $\tz_{\rm p}>\tL/2$ where the steric penalty disappears, one has $S_{\rm p}(\tz_{\rm p})>0$, i.e. the polymer exhibits a weak tendency to orient itself perpendicular to the membrane. The origin of this orientational transition will be investigated in Section~\ref{1lsalt}.

In Fig.~\ref{fig2}(b), the strong effect of the tetravalent counterions as the glue of the like-charged polymer adsorption can be realized by noting that despite the weak membrane charge $\sigma_{\rm m}=0.2$ $e/\rm{nm}^2$, added counterions of milimolar concentration rises the polymer density by several factors above its bulk value. For a better insight into this effect, we consider the large distance regime $\tz_{\rm p}\gg1$ where Eq.~(\ref{eqs19}) takes the asymptotic form
\bea\label{eqs21}
\beta\Delta\Omega^{(0)}_{\rm pc}(z_\p,\te)&\approx&-4\gamma\Gamma_{\rm c}L_{\rm p}(\te)\tau e^{-\tz_{\rm p}}\\
&&\hspace{-5mm}\times\left\{\tz_{\rm p}+\frac{3}{2}-\frac{J_1(0)}{4\gamma\qc}-\frac{\tL\ct/2}{\tanh\left(\tL\ct/2\right)}\right\}.\nonumber
\eea
Moreover, we note that at large distances $\tz_\p\gg1$, the polymer-membrane interaction energy~(\ref{11}) simplifies to 
\be\label{sas4II}
\beta\Omega_{\rm pm}^{(0)}(z_\p,\te)\approx4\gamma L_\p(\te)\tau e^{-\tz_\p}.
\ee
One sees that due to the polymer \RP{location} inside the bracket of Eq.~(\ref{eqs21}), the attractive polymer-counterion coupling potential is longer ranged than the repulsive polymer-membrane coupling energy~(\ref{sas4II}). Thus, in the presence of a substantial amount of multivalent counterions, weakly charged polymers located at large separation distances will always feel an attraction by the like-charged membrane surface.

\subsubsection{Effect of the membrane charge magnitude on the like-charge attraction}

In order to better understand the physical mechanism behind the like-charge polymer adsorption, we focus now on the effect of the membrane charge \RP{magnitude}. To this end, by using the asymptotic laws~(\ref{eqs21}) and~(\ref{sas4II}), we recast the large distance limit of the grand potential~(\ref{eqs16}) in a form similar to the DH interaction potential~(\ref{12}),
\be\label{eqs23}
\beta\Delta\Omega^{(0)}_{\rm p}(z_\p,\te)\approx\frac{2}{s}\tau L_{\rm p}(\te)\eta_{\rm c}(\tz_{\rm p},\te)e^{-\tz_{\rm p}},
\ee
where we introduced the auxiliary function
\bea\label{eqs24}
&&\eta_{\rm c}(\tz_{\rm p},\te)=\\
&&2s\gamma\left\{1-\Gamma_{\rm c}\left[\tz_{\rm p}+\frac{3}{2}-\frac{J_1(0)}{4\gamma\qc}-\frac{\tL\ct/2}{\tanh\left(\tL\ct/2\right)}\right]\right\}.\nonumber
\eea
Eq.~(\ref{eqs24}) is a  non-uniform charge renormalization function dressed by MF-level non-linearities and the charge correlations originating from the multivalent counterions. 

Figs.~\ref{fig3}(a) and (b) display for $\theta_\p=\pi/2$ the effect of the membrane charge magnitude in terms of the potential profiles~(\ref{eqs16}) and~(\ref{eqs19}) (curves), and their asymptotic limits in Eqs.~(\ref{eqs21}) and~(\ref{eqs23}) (symbols). In neutral membranes where the repulsive MF potential~(\ref{sas4II}) vanishes (black curves), the polymer grand potential tends to the polymer-counterion coupling energy
\be\label{eqs24II}
\lim_{\sigma_{\rm m}\to0}\Delta\Omega^{(0)}_{\rm p}(z_\p,\te)=\Delta\Omega^{(0)}_{\rm pc}(z_\p,\te)=\frac{\Gamma_{\rm c}}{\beta q_{\rm c}}\tau L_{\rm p}(\te)e^{-\tz_{\rm p}}.
\ee
The weakly repulsive energy~(\ref{eqs24II}) originates from a multivalent cation-induced \RP{effect, akin to the "image-charge" interactions, with an origin in the confinement of the counterions and salt to the $z \geq 0$ region}; the screening of the polymer charges  in the bulk solution including the multivalent counterions is more efficient than in the region close to the counterion-free membrane. This translates into a repulsive force driving the polymer away from the membrane surface.

\begin{figure*}
\includegraphics[width=1\linewidth]{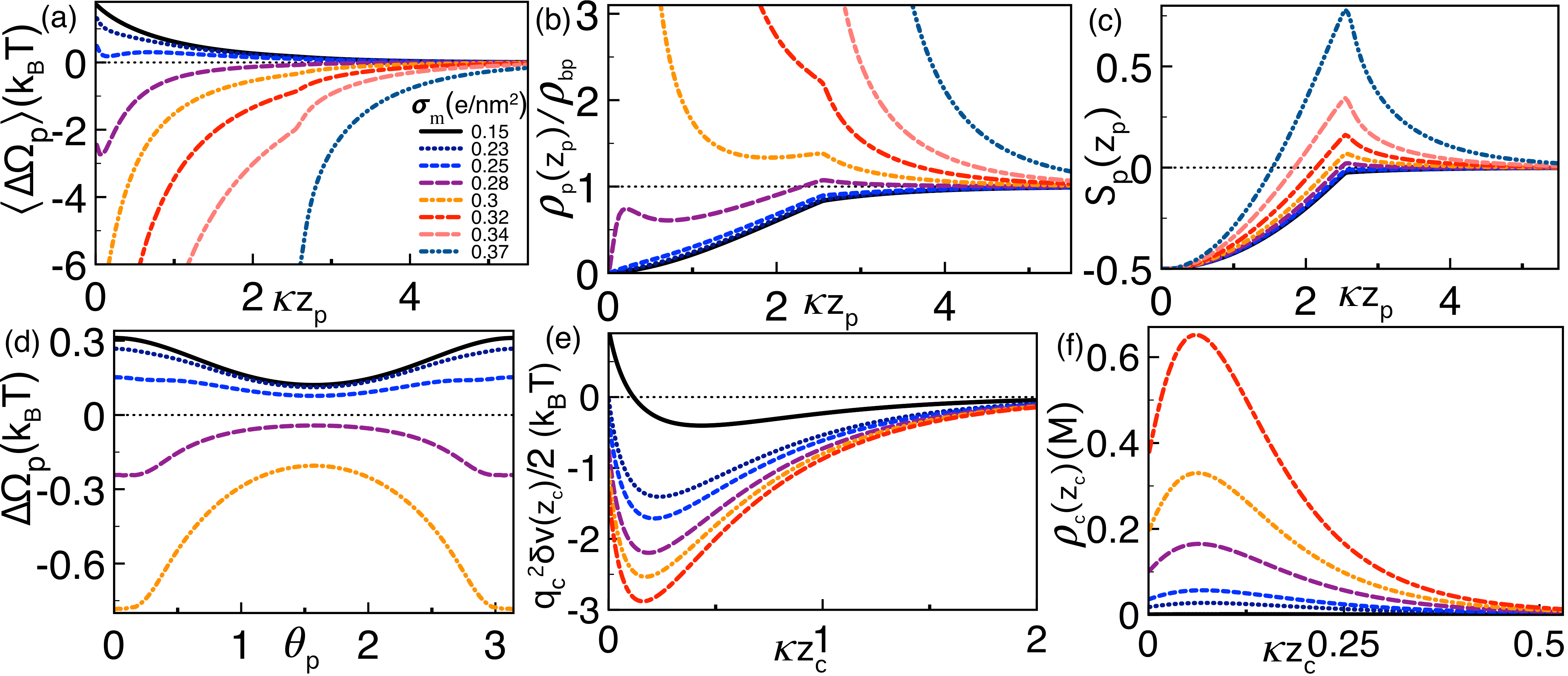}
\caption{(Color online) (a) Grand potential~(\ref{netpot}) averaged over polymer rotations, (b) polymer density~(\ref{dn}), (c) \RP{\sl orientational order parameter}~(\ref{or}), (d) the angular dependence of the polymer grand potential at $z_\p=L/2$, (e) ionic self-energy contribution to the counterion density~(\ref{eqs11}) obtained from Eq.~(\ref{apeq25}), and (f) counterion density profile. The polymer length and charge are $L=5.0$ nm and $\bt=0.05$. The monovalent salt and tetravalent counterion densities are $\rho_{\rm b}=0.1$ M and $\rho_{\rm bc}=0.1$ mM.}
\label{fig4}
\end{figure*}

In the opposite case of a charged membrane, the excess of the multivalent counterions attracted by the surface reverses this balance; the interfacial liquid able to screen the polymer charges more efficiently than the bulk solution favours the location of the molecule close to the substrate. In Figs.~\ref{fig3}(a) and (b), this effect manifests itself by the switching of the interaction potentials from repulsive to attractive, and the amplification of the polymer binding with increasing membrane charge, i.e. $\sigma_{\rm m}\uparrow\Delta\Omega^{(0)}_{\rm pc}\downarrow\Delta\Omega^{(0)}_{\rm p}\downarrow$. The intensification of the like-charge polymer-membrane complexation with the membrane charge magnitude has been indeed observed in the adsorption experiments of Ref.~\cite{Caball2014} where the dipalmitoylphosphatidyslerine-rich regions of the anionic substrate characterized by a higher charge density were found to be occupied by larger amounts of DNA aggregates.

Fig.~\ref{fig3}(b) indicates that the increase of the membrane charge \RP{moves} the grand potential minimum closer to the interface. The corresponding binding position of the polymer follows from the equality $\partial_{z_{\rm p}}\left.\Delta\Omega^{(0)}_{\rm p}(z_\p,\te)\right |_{\tz^*_{\rm p}}=0$ and Eq.~(\ref{eqs23}) as
\be\label{eqs25}
\tz^*_{\rm p}=\frac{1}{\Gamma_{\rm c}}-\frac{1}{2}+\frac{J_1(0)}{4\gamma\qc}+\frac{\tL\ct/2}{\tanh\left(\tL\ct/2\right)}.
\ee
Fig.~\ref{fig3}(c) illustrates the membrane charge dependence of the position $\tz^*_{\rm p}$ obtained numerically from the grand potential~(\ref{eqs16}) (curves) and the formula~(\ref{eqs25}) (symbols). One sees that due to the enhanced counterion attraction, the increment of the membrane charge or counterion concentration \RP{drives} the like-charged polymer to the surface, i.e. $\sigma_{\rm m}\uparrow\tz^*_{\rm p}\downarrow$ and $\rho_{\rm bc}\uparrow\tz^*_{\rm p}\downarrow$. Indeed, Eq.~(\ref{eqs25}) indicates that in the weak membrane charge regime of Fig.~\ref{fig3}(c) where $s\gtrsim1$, the binding position drops according to an inverse linear function of the membrane charge density and counterion valency, i.e. $\tz^*_{\rm p}\approx s/(2q_{\rm c})\propto\left(q_{\rm c}\sigma_{\rm m}\right)^{-1}$. Moreover, due to the first term of Eq.~(\ref{eqs25}), the binding position decreases as well as an inverse linear function of the counterion concentration  $\rho_{\rm bc}$. In Sec.~\ref{1lsalt}, we investigate the alterations in the polymer adsorption mechanism 
by salt correlations emerging beyond the present weak charge regime.

\subsection{Cooperative effect of correlations by mono- and multivalent ions : \RP{$1\ell$} salt}
\label{1lsalt}

\subsubsection{Emergence of \RP{$1\ell$} salt correlations: weakly anionic polymers}
\label{shpol}

We consider here the departure from the  MF salt regime of Sec.~\ref{mfsalt}. To this end, we focus on intermediate membrane charge magnitudes and investigate the impact of the emerging salt correlations on the adsorption of weakly charged polyelectrolytes. Figs.~\ref{fig4}(a) and (b) display the orientation-averaged polymer grand potential~(\ref{netpot}) and density~(\ref{dn}) including \RP{$1\ell$}-level salt correlations at the low tetravalent counterion  concentration $\rho_{\rm bc}=0.1$ mM. At weakly charged membranes with surface charge $\sigma_{\rm m}=0.15$ $\mbox{e}/\mbox{nm}^2$, MF-level repulsive interactions ($\Delta\Omega_{\rm p}>0$) result in the interfacial exclusion of the polymer, i.e. $\rho_{\rm p}(z_\p)<\rho_{\rm bp}$ (black curves). Then, upon the slight increase of the membrane charge, the grand potential for $\sigma_{\rm m}>0.25$ $\mbox{e}/\mbox{nm}^2$ turns from positive to strongly negative. The corresponding attractive force on the polymer rises the interfacial polymer density by orders of magnitude ($\sigma_{\rm m}\uparrow\rho_{\rm p}\uparrow$) and also results in a thick adsorption layer extending over several DH lengths.

We focus now on the orientational configuration of the polymer. Figs.~\ref{fig4}(c) and (d) display the \RP{\sl orientational order parameter}~(\ref{or}) and the angular dependence of the polymer grand potential. In the case of weak membrane charges where the grand potential is minimized at $\te=\pi/2$ (black curve), the underlying like-charge polymer-membrane repulsion results in the parallel orientation of the polymer with the membrane, i.e. $S_{\rm p}<0$. Then, the rise of the membrane charge into the regime $\sigma_{\rm m}>0.25$ $\mbox{e}/\mbox{nm}^2$  turns the grand potential $\Delta\Omega_{\rm p}$ from concave to convex and increases the \RP{\sl orientational order parameter} to strongly positive values $S_{\rm p}>0$, i.e. $\sigma_{\rm m}\uparrow S_{\rm p}\uparrow$. Hence, beyond a characteristic charge magnitude, the like-charge polymer attraction is accompanied with the orientational transition of the molecule from parallel to perpendicular configuration. This behaviour is similar to the reorientation of dipolar molecules by the increase of the electrostatic coupling parameter~\cite{Podgornik2009}.
\begin{figure}
	\includegraphics[width=1.\linewidth]{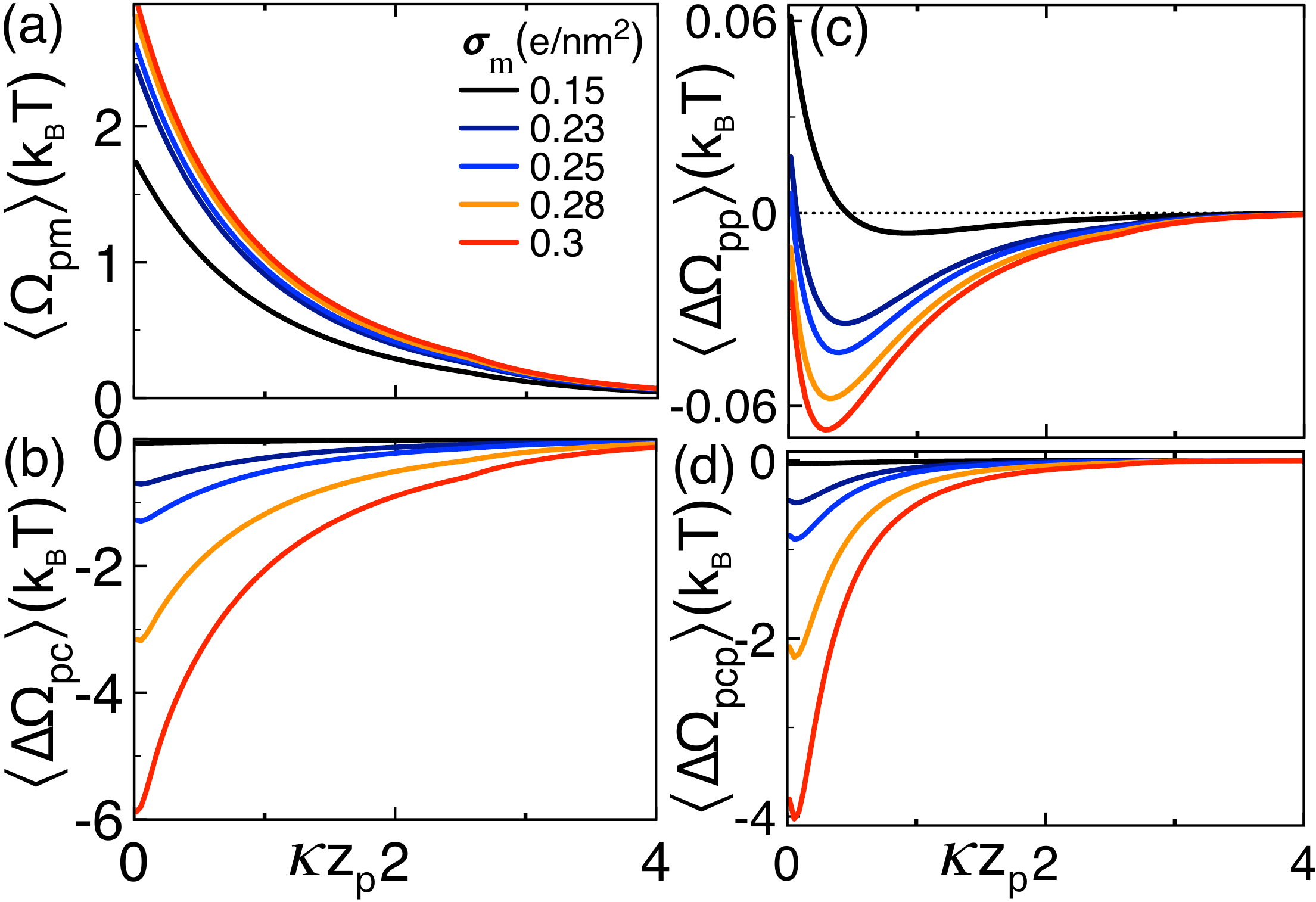}
	\caption{(Color online) (a) Polymer-membrane interaction potential~(\ref{eqs12}), (b) polymer-counterion coupling energy~(\ref{eqs13}), (c) polymer self-energy~(\ref{eqs14}), and (d) screening energy of the polymer self-interaction~(\ref{eqs15}) averaged over polymer rotations. The model parameters are the same as in Fig.~\ref{fig4}.}
	\label{fig5}
\end{figure}

It should be noted that in the present intermediate membrane charge regime, the counterion concentration $\rho_{bc}=0.1$ mM for the occurrence of the like-charge adsorption is more than two orders of magnitude lower than in the weak membrane charge regime considered in Section~\ref{mfsalt}. Moreover, the comparison of Figs.~\ref{fig2}(b) (inset) and~\ref{fig4}(c) indicates that despite the significantly lower counterion concentration, the perpendicular polymer orientation is considerably stronger than in the weak membrane charge regime. These peculiarities originate from the emergence of salt correlations with the increment of the membrane charge. Namely, the monovalent counterion excess at the charged surface enhances the screening ability of the interfacial liquid. In Fig.~\ref{fig4}(e), we show that the enhanced interfacial screening gives rise to an attractive ionic self-energy intensified with the membrane charge magnitude, i.e. $\sigma_{\rm m}\uparrow\delta v (z_\p)\downarrow$. Fig.~\ref{fig4}(f) indicates that this additional attraction brings further tetravalent counterions to the surface and amplifies the average counterion density~(\ref{eqs11}) by more than three orders of magnitude above its bulk value, i.e. $\sigma_{\rm m}\uparrow\rho_{\rm c}\uparrow$. Thus, for weakly charged polymers, the main effect of salt correlations emerging at intermediate membrane charges consists of enhancing the adhesive force of the counterions bridging the \RP{space between} the polymer and the like-charged membrane.

In order to gain further insight into the impact of salt correlations on polymer-membrane interactions, in Fig.~\ref{fig5}, we plotted the grand potential components in Eqs.~(\ref{eqs12})-(\ref{eqs15}) from weak to intermediate membrane charge coupling. It should be first noted that as one increases the membrane charge beyond the MF salt regime, the amplification of the \RP{$1\ell$} potential correction $\phi_{\rm m}^{(1)}>0$ of Eq.~(\ref{potsup1}) opposing the negative MF potential component $\phi_{\rm m}^{(0)}<0$ attenuates the rise of the polymer-membrane coupling energy $\Omega_{\rm pm}$ in Eq.~(\ref{eqs12}). Consequently,  in Fig.~\ref{fig5}(a), the net \RP{$1\ell$}-level repulsion energy $\Omega_{\rm pm}$ saturates at intermediate membrane charges.

In the same membrane charge regime, Figs.~\ref{fig5}(a)-(d) show that the relevant grand potential components driving the adsorption transition are the repulsive polymer-membrane interaction energy $\Omega_{\rm pm}$, and the counterion-induced attractive components $\Delta\Omega_{\rm pc}$ and $\Delta\Omega_{\rm pcp}$ of comparable magnitude. That is, the polymer self-energy $\Delta\Omega_{\rm pp}$ resulting solely from salt correlations is perturbative. Moreover, the screening energy $\Delta\Omega_{\rm pcp}$ is shorter ranged than the other potential components and becomes perturbative in the region $\tz_{\rm p}\gtrsim2$ where the orientational transition of the polymer occurs (see Fig.~\ref{fig4}(c)). Thus, while enhancing the polymer binding very close to the membrane surface, the screening energy $\Delta\Omega_{\rm pcp}$  brings a minor contribution to the orientational transition of the molecule mainly driven by the polymer-counterion coupling energy $\Delta\Omega_{\rm pc}$.

\begin{figure}
\includegraphics[width=.9\linewidth]{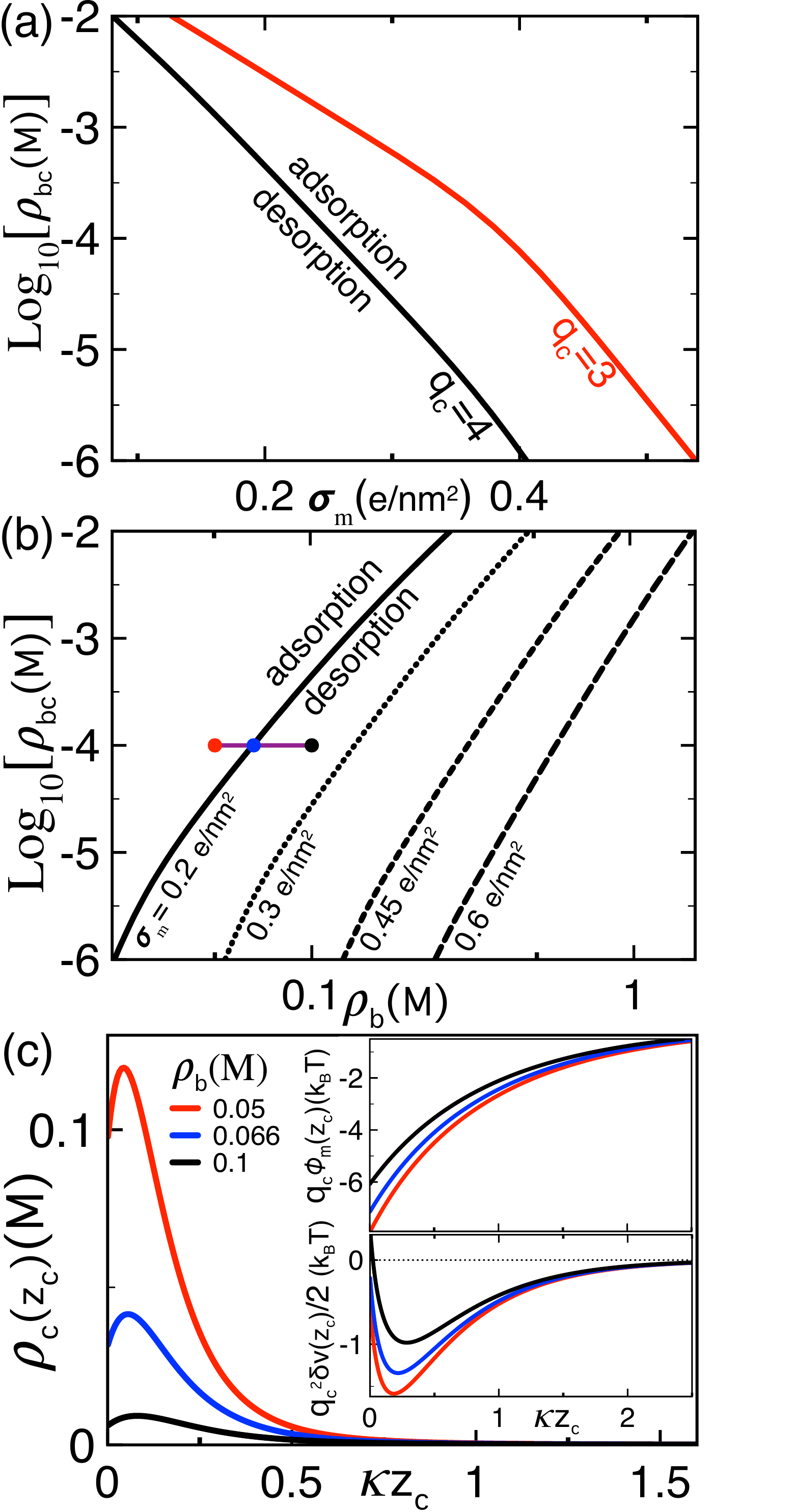}
\caption{(Color online) Phase diagrams: critical cation concentration for polymer binding (a) against the membrane charge at the salt concentration $\rho_{\rm b}=0.1$ M, and (b) versus the salt concentration ($q_{\rm c}=4$). (c) Tetravalent counterion density (main plot), and the average potential and ionic self-energy contributions in Eq.~(\ref{eqs11}) (inset). In (c), the membrane charge is $\sigma_{\rm m}=0.2$ $\mbox{e}/\mbox{nm}^2$, counterion concentration $\rho_{\rm bc}=0.1$ mM, and the salt density  $\rho_{\rm b}$ for each curve corresponds to the salt density value of the dot with the same color in (b). The other model parameters are the same as in Fig.~\ref{fig4}.}
\label{fig6}
\end{figure}
\begin{figure*}
\includegraphics[width=1.\linewidth]{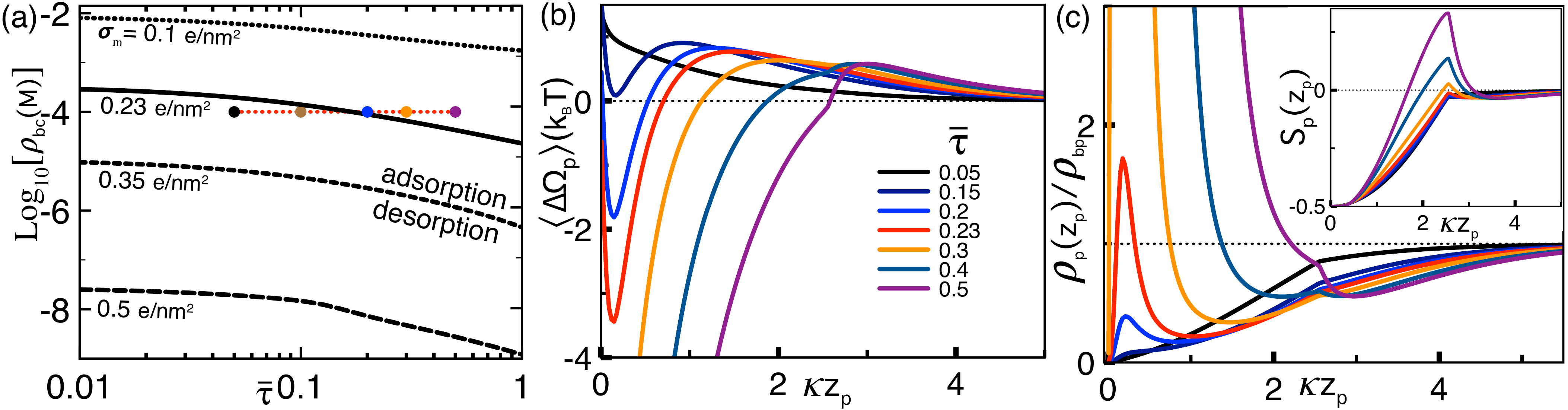}
\caption{(Color online) (a) Phase diagram: critical tetravalent cation concentration against the dimensionless polymer charge $\bt$ at the salt concentration $\rho_{\rm b}=0.1$ M. (b) Grand potential~(\ref{netpot}) averaged over polymer rotations, and (c) polymer density~(\ref{dn}) (main plot) and \RP{\sl orientational order parameter}~(\ref{or}) (inset) at the counterion density $\rho_{\rm bc}=10^{-4}$ M and membrane charge $\sigma_{\rm m}=0.23$ $\mbox{e}/\mbox{nm}^2$. The polymer charge for each curve is given in the legend of (b). The other model parameters are the same as in Fig.~\ref{fig4}.}
\label{fig7}
\end{figure*}

To summarize, beyond the weak membrane charge regime, the growth of salt correlations amplifying the interfacial counterion excess results in the emergence of the attractive screening energy  $\Delta\Omega_{\rm pcp}$ and the amplification of the polymer-counterion interaction energy $\Delta\Omega_{\rm pc}$. These two effects are responsible for the enhancement of the like-charge polymer adsorption by leading order salt correlations at intermediate membrane charge magnitudes. The impact of this mechanism on the critical adsorption point is illustrated in the phase diagram of Fig.~\ref{fig6}(a). The critical lines mark the characteristic counterion concentration $\rho^*_{\rm bc}$ where the average grand potential in Fig.~\ref{fig4}(a) switches at $z_{\rm p}=0$ from positive to negative. Fig.~\ref{fig6}(a) shows that with increasing membrane charge, the critical counterion density line separating the polymer adsorption and desportion regimes drops quickly by orders of magnitude, i.e. $\sigma_{\rm m}\uparrow\rho^*_{\rm bc}\downarrow$. One also notes that due to the rise of the coupling parameter~(\ref{eqs22}), the larger the counterion valency, the lower the critical counterion density at the adsorption transition, i.e. $q_{\rm c}\uparrow\rho^*_{\rm bc}\downarrow$. This trend is in agreement with the adsorption experiments of Ref.~\cite{Tiraferri2015} where the critical counterion density maximizing the like-charge polymer binding was observed to drop with the increase of the counterion valency.

Finally, the phase diagram in Fig.~\ref{fig6}(b) illustrates the effect of salt on the critical counterion density. One sees that at fixed membrane charge, a weak increment of the salt density in the submolar regime rises the critical  counterion concentration by several orders of magnitude, i.e. $\rho_{\rm b}\uparrow\rho^*_{\rm bc}\uparrow$. Moreover,  adding salt and crossing horizontally one of the critical lines at fixed counterion density (e.g. via the purple line), the system switches from polymer adsorption to desorption state. One also notes that the rise of the membrane charge moves the critical line towards larger salt concentrations. Hence, the minimum membrane charge $\sigma_{\rm m}^*$ for polymer adsorption increases with the amount of salt, i.e. $\rho_{\rm b}\uparrow\sigma_{\rm m}^*\uparrow$. These points indicate that  added salt causes the unbinding of the polymer from the membrane. This peculiarity has been equally observed in the experiments of Ref.~\cite{Tiraferri2015} and the simulations of Ref.~\cite{Levin2016} where the addition of monovalent salt was found to result in the decomplexation of the polymer and the like-charged substrate.

Fig.~\ref{fig6}(c) shows that polymer desorption by salt addition originates from the suppression of attractive charge correlations at two different levels in Eq.~(\ref{eqs11}). First, salt ions screen the average membrane potential, i.e. $\rho_{\rm b}\uparrow|\phi_{\rm m}|\downarrow$  (see the top plot of the inset). This weakens the direct multivalent counterion attraction by the membrane charges. Then, via the screening of the monovalent cation attraction to the membrane surface,  added salt reduces as well the interfacial monovalent cation excess, and lowers the magnitude of the attractive 
energy originating from this excess, i.e. $\rho_{\rm b}\uparrow|\delta v|\downarrow$ (see the bottom plot in the inset). As a result of both effects, salt addition strongly suppresses the interfacial counterion density, $\rho_{\rm b}\uparrow\rho_{\rm c}(z_{\rm c})\downarrow$ (see the main plot). This weakens the net adhesive force of the counterions mediating the like-charge polymer binding and leads to the desorption of the polymer from the substrate.

\subsubsection{Adsorption of strongly anionic polymers}
\label{lgpol}

The polymer charge magnitude enhances both the repulsive polymer-membrane coupling energy in Fig.~\ref{fig5}(a), and the opposing attractive interaction components in Fig.~\ref{fig5}(b)-(d). In order to understand the net effect of the polymer charge density on the adsorption of the molecule, we relax now the weakly charged polymer assumption. In Fig.~\ref{fig7}(a), we \RP{display} the evolution of the critical counterion density with the polymer charge between $\bt=0.01$ and the dsDNA value $\bt=1.0$. One first notes that the rise of the polymer charge density at fixed counterion concentration switches the system from polymer desorption to adsorption state. One also sees that the at fixed membrane charge magnitude $\sigma_{\rm m}$, the higher the polymer charge density, the lower the critical counterion concentration for polymer adsorption, i.e. $\bt\uparrow\rho^*_{\rm bc}\downarrow$. We finally note that the increment of the membrane charge density drops the critical line towards lower counterion concentration regimes. As a result, at fixed counterion concentration, the higher the membrane charge density, the weaker the minimum polymer charge $\bt^*$ for the occurrence of the like-charge adsorption, i.e. $\sigma_{\rm m}\uparrow\bt^*\downarrow$.

These trends indicate that the net effect of the polymer charge magnitude is the amplification of charge correlations. In Figs.~\ref{fig7}(b) and (c), this point is illustrated in terms of the polymer grand potential and density, and the \RP{\sl orientational order parameter}. Starting at the black circle of Fig.~\ref{fig7}(a) where the polymer is unbound, and crossing the critical line by rising the molecular charge $\bt$ via the red curve, the grand potential turns from repulsive to strongly attractive, i.e. $\bt\uparrow\Delta\Omega_{\rm p}\downarrow$. Consequently, the polymer density develops an adsorption peak that grows together with the \RP{\sl orientational order parameter}, i.e. $\bt\uparrow\rho_{\rm p}\uparrow S_{\rm p}\uparrow$. Thus, in the presence of multivalent counterions, the increment of the polymer charge magnitude can solely trigger the orientational transition of the polymer and the subsequent like-charge adsorption of the molecule.

\begin{figure}
\includegraphics[width=1.\linewidth]{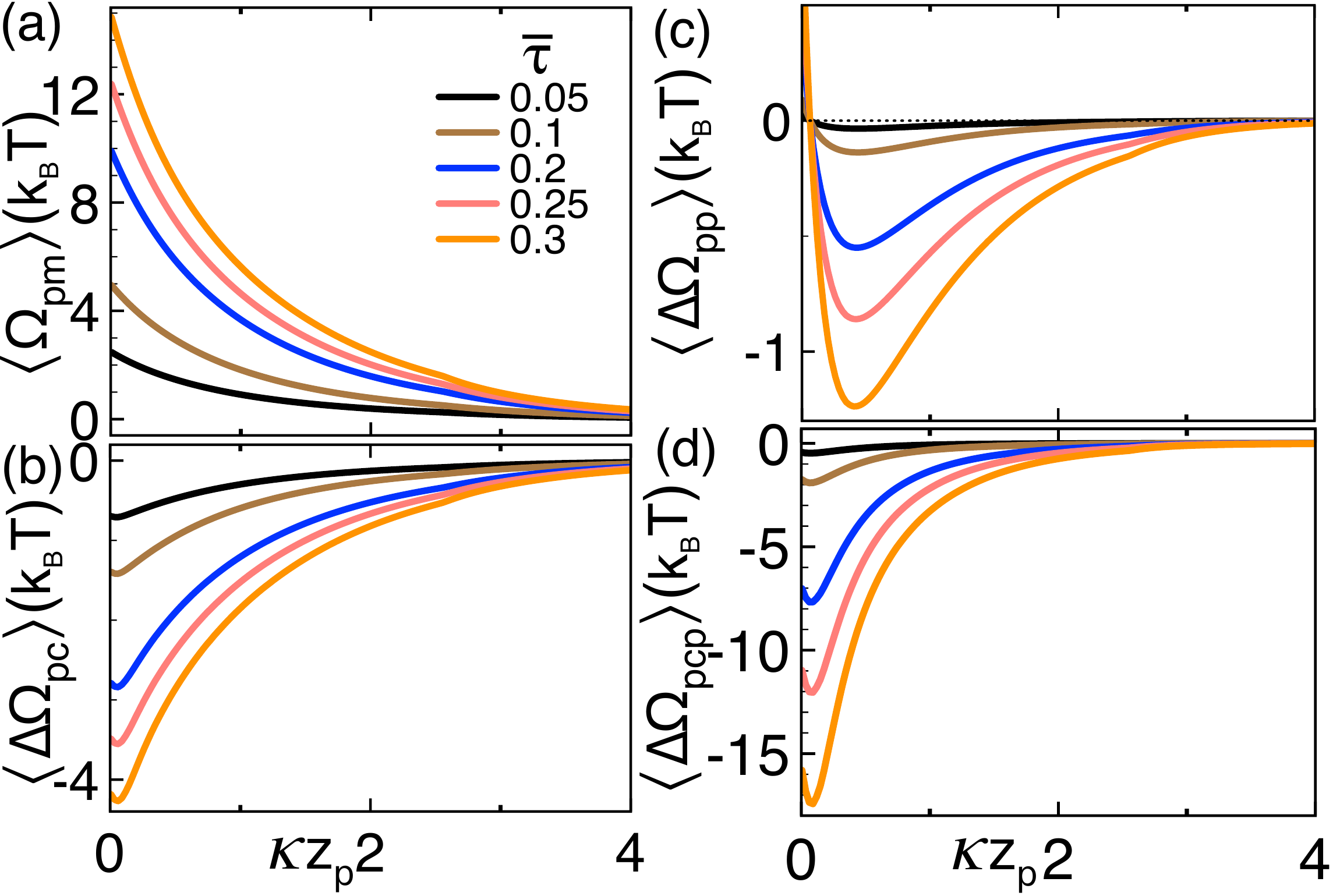}
\caption{(Color online) (a) Polymer-membrane interaction potential~(\ref{eqs12}), (b) polymer-counterion coupling energy~(\ref{eqs13}), (c) polymer self-energy~(\ref{eqs14}), and (d) screening energy of the polymer self-interaction~(\ref{eqs15}) averaged over polymer rotations. The membrane charge is $\sigma_{\rm m}=0.23$ ${\rm e}/{\rm nm}^2$. The polymer charge density for each curve is given in the legend of (a). The other model parameters are the same as in Fig.~\ref{fig7}.}
\label{fig8}
\end{figure}

In order to identify the specific mechanism driving the adsorption of strongly charged polymers, in Fig.~\ref{fig8}, we \RP{display} the grand potential components~(\ref{eqs12})-(\ref{eqs15}) at the polymer charge densities corresponding to the dots of the same color in Fig.~\ref{fig7}(a). First of all, one sees that the polymer self-energy of small magnitude $|\Delta\Omega_{\rm pp}|\lesssim k_{\rm B}T$ brings a perturbative contribution to like-charge polymer binding. Furthermore, as one gradually increases the polymer charge $\bt$ via the red curve in Fig.~\ref{fig7}(a), the attractive screening energy $|\Delta\Omega_{\rm pcp}|$ exceeds the polymer-counterion coupling potential $|\Delta\Omega_{\rm pc}|$ at $\bt\approx0.1$ (brown curves in Fig.~\ref{fig8}(b) and (d)). Subsequently, the like-charge attraction occurs at the polymer charge magnitude $\bt\approx0.2$ where the screening energy $\Delta\Omega_{\rm pcp}$ is twice as large in magnitude as the potential component $\Delta\Omega_{\rm pc}$ (blue curves). Thus, at the transition point, the system is mainly governed by the competition between the repulsive polymer-membrane coupling potential $\Omega_{\rm pm}$ and the attractive screening energy $\Delta\Omega_{\rm pcp}$. 

The dominant effect of the screening energy $\Delta\Omega_{\rm pcp}$ can be also noted by increasing further the polymer charge to the value $\bt=0.3$ where the attractive potential $\Delta\Omega_{\rm pcp}$ solely takes over the repulsive coupling potential $\Omega_{\rm pm}$ close to the interface (orange curves in Fig.~\ref{fig8}(a) and (d)). This indicates that the adsorption of strongly charged polymers such as ssDNA ($\bt=0.5$) and dsDNA molecules ($\bt=1.0$) is essentially driven by the short ranged energy $\Delta\Omega_{\rm pcp}$. In Fig.~\ref{fig7}(c), the \RP{non-monotonic behavior}  of the polymer density originates indeed from this peculiarity. Namely, Fig.~\ref{fig7}(b) indicates that as the repulsive potential $\Omega_{\rm pm}$ of longer range dominates the screening energy $\Delta\Omega_{\rm pcp}$ far from the interface, regardless of the charge magnitude $\bt$, the grand potential always keeps a repulsive branch outside the interfacial region. As a result, Fig.~\ref{fig7}(c) shows that in the case of strongly charged polymers, the interfacial adsorption peak of the polymer density is always accompanied with a polymer depletion layer at larger distances from the membrane surface.

\section{Conclusions}

The characterization of the \RP{strong coupling electrostatic forces} mediated by multivalent counterions is essential for understanding and controlling {\sl in vivo} and {\sl in vitro} biological processes involving charged macromolecules.  In this article, we probed the physical mechanisms behind the orientational transition and the subsequent adsorption of short polyelectrolytes onto like-charge membranes by multivalent counterion addition into a monovalent salt solution. In order to shed light on the nature of the adhesive forces induced by the multivalent counterions on the polymer-membrane complex, we developed \RP{a statistical mechanical formalism} that can take into account charge correlations associated with the monovalent salt at \RP{$1\ell$}-level, and the presence of multivalent cations at SC-level. We summarize below our main results and conclusions.

Within the framework of our \textit{\RP{$1\ell$}-dressed SC theory}, we found that the effect of the multivalent counterions bridging the polyelectrolyte and the like-charge membrane originates in their interfacial excess in the vicinity of the membrane surface. This counterion excess locally maximizes the screening ability of the electrolyte close to the interface and minimizes the electrostatic grand potential of the polyelectrolyte. This translates into an effective force driving the polymer towards the substrate. 

The details of this mechanism were scrutinized in different regimes of the charge magnitude. In the case of weakly charged polymers and membranes where salt ions behave at the MF-level, the adsorption transition is driven by the competition between the repulsive polymer-membrane coupling potential $\Omega_{\rm pm}$ and the attractive polymer-counterion interaction potential $\Delta\Omega_{\rm pc}$. Then, the increment of the membrane charge beyond the WC regime results in the emergence of salt correlations. First, these correlations attenuate the growth of the repulsive energy $\Omega_{\rm pm}$ with the membrane charge magnitude, \RP{and second}, they give rise to an attractive ionic self-energy $\delta v$ that significantly enhances the multivalent counterion excess at the membrane surface. The enhanced interfacial counterion density results in the amplification of the polymer-counterion coupling energy $\Delta\Omega_{\rm pc}$ and the emergence of the additional attractive screening energy $\Delta\Omega_{\rm pcp}$ of comparable magnitude. Due to the combination of these effects, salt correlations growing at intermediate membrane charges reinforce the polymer-membrane complex and lowers the critical counterion concentration for like-charge complexation by orders of magnitude, i.e. $\sigma_{\rm m}\uparrow\rho_{\rm bc}^*\downarrow$.

We also found that upon the rise of the dimensionless polymer charge density beyond the WC regime $\bt\gtrsim0.1$, the screening energy $\Delta\Omega_{\rm pcp}$ takes over the polymer-counterion coupling potential $\Delta\Omega_{\rm pc}$ and becomes the dominant attractive potential component at play. This indicates that the adsorption of strongly charged biopolymers such as DNA molecules is driven by the competition between the repulsive polymer-membrane coupling potential $\Omega_{\rm pm}$ and the attractive screening energy $\Delta\Omega_{\rm pcp}$. 

Finally, we showed that the rise of the counterion valency amplifies charge correlations and lowers the critical multivalent counterion concentration for polymer adsorption, i.e. $q_{\rm c}\uparrow\rho_{\rm bc}^*\downarrow$. In addition, added monovalent salt was found to suppress charge correlations and to reduce the interfacial counterion density, leading to the desorption of the polyelectrolyte from the like-charged membrane. We emphasize that the reduction of the critical counterion concentration with increasing ion valency and the salt-induced unbinding of the polymer from the membrane have been observed in adsorption experiments~\cite{Tiraferri2015}. 

Our formalism includes certain approximations that can be relaxed in future works. For example, our polyelectrolyte model is based on the stiff polymer approximation reasonable for the short polymers considered \RP{above}. In order to characterize the adsorption of long DNA sequences, the conformational polymer fluctuations can be incorporated into our model  within the field theoretic formalism that treats the ions and the polymer charges on the same footing~\RP{\cite{Podgornikpoly, dun}}. Moreover, the derivation of the \RP{$1\ell$}-dressed SC theory is based on the low fugacity expansion of the electrostatic grand potential in terms of the multivalent counterion density. This approximation considers the multivalent cations as test charges and therefore neglects their effect on the ionic environment. With the aim to improve over this approximation, we are currently working on a self-consistent formulation of the \RP{$1\ell$}-dressed SC theory. We also plan to confront in a future work the predictions of our theory with numerical simulations. This will allow us to determine quantitatively the validity regime of our approximations. We finally note that our detailed phase diagrams in Figs.~\ref{fig6}(a)-(b) and~\ref{fig7}(a) can provide valuable guiding information for future adsorption experiments.

\smallskip
\appendix

\begin{widetext}
\section{Auxiliary function $J_n(x)$ in Eq.~(\ref{eqs20})}
\label{apI}
We report here the auxiliary function $J_n(x)$ defined in Eq.~(\ref{eqs20}). For trivalent counterions $q_{\rm c}=3$, one has
\bea\label{apeq1}
J_{-1}(x)&=&-e^{-x}-\frac{12}{\gamma}\ln\left(1-\gamma e^{-x}\right)+\frac{4}{5\gamma}\frac{e^{x}}{\left(e^x-\gamma\right)^5}\left\{-31e^{4x}+140\gamma e^{3x}-250\gamma^2e^{2x}+200\gamma^3e^{x}-75\gamma^4\right\},\\
\label{apeq2}
J_0(x)&=&x-\frac{4\gamma}{15\left(e^x-\gamma\right)^5}\left\{45e^{4x}-90\gamma e^{3x}+140\gamma^2e^{2x}-70\gamma^3e^x+23\gamma^4\right\},\\
\label{apeq3}
J_1(x)&=&12\gamma\ln\left(e^{x}-\gamma\right)\\
&&+\frac{1}{5\left(e^x-\gamma\right)^5}\left\{5e^{6x}-25\gamma e^{5x}-250\gamma^2e^{4x}+750\gamma^3e^{3x}-975\gamma^4e^{2x}+555\gamma^5e^x-124\gamma^6\right\},\nonumber
\eea
with $J_{-1}(\infty)=-124/(5\gamma)$. For tetravalent counterions $q_{\rm c}=4$, the integral in Eq.~(\ref{eqs20}) yields
\bea
\label{apeq4}
J_{-1}(x)&=&-e^{-x}-\frac{16}{\gamma}\ln\left(1-\gamma e^{-x}\right)\\
&&+\frac{16e^x}{105\gamma\left(e^x-\gamma\right)^7}\left\{-247e^{6x}+1624\gamma e^{5x}-4557\gamma^2e^{4x}+6860\gamma^3e^{3x}-6125\gamma^4e^{2x}+2940\gamma^5e^x-735\gamma^6\right\},\nonumber\\
\label{apeq5}
J_0(x)&=&x-\frac{16\gamma}{105\left(e^x-\gamma\right)^7}\left\{105e^{6x}-315\gamma e^{5x}+770\left[\gamma^2e^{4x}-\gamma^3e^{3x}\right]+609\gamma^4e^{2x}-203\gamma^5e^x+44\gamma^6\right\},\\
\label{apeq6}
J_1(x)&=&\frac{1}{105\left(e^x-\gamma\right)^7}\left\{105e^{8x}-735\gamma e^{7x}-9555\gamma^2e^{6x}+43365\gamma^3 e^{5x}-94325\gamma^4e^{4x}+107555\gamma^5e^{3x}\right.\\
&&\left.\hspace{2.2cm}-72177\gamma^6e^{2x}+25879\gamma^7e^x-3952\gamma^8+1680\gamma\left(e^x-\gamma\right)^7\ln\left(e^x-\gamma\right)\right\},\nonumber
\eea
with $J_{-1}(\infty)=-3952/(105\gamma)$. 

\end{widetext}

\section{Computation of the \RP{$1\ell$} ionic potentials}
\label{pots1l}

We explain here the derivation of the correlation corrections to the average potential and correlator in Eqs.~(\ref{potsup1}) and~(\ref{potsup2}). The details of the calculation summarized here can be found in Ref.~\cite{Buyuk2012}. 

\subsection{Computing the Green's function and ionic self-energy}

According to the \RP{$1\ell$} theory of inhomogeneously distributed monovalent salt solutions, the electrostatic Green's function solves the non-uniformly screened Green's equation
\be
\label{apeq15}
\nabla^2 v(\br,\br')-\kappa^2_{\rm c}(\br)v(\br,\br')=-4\pi\ell_{\rm B}\delta(\br-\br'),
\ee
with the local charge screening function 
\be\label{apeq16}
\kappa_{\rm c}^2(\br)=\kappa^2\cosh\left[\phi_0(\br)\right], 
\ee
where the MF-level average potential $\phi_0(\br)$ solves the PB equation~(\ref{PB}). Considering now the planar symmetry and using the Fourier expansion~(\ref{grsym}), Eq.~(\ref{apeq15}) becomes
\bea
\label{apeq17}
\left[\de^2-p_{\rm c}^2(z)\right]\tv(z,z')=-4\pi\ell_{\rm B}\delta(z-z'),
\eea
with the auxiliary screening function
\be
\label{apeq18}
p_{\rm c}^2(z)=\kappa_{\rm c}^2(z)+k^2.
\ee

For the single interface system depicted in Fig.~\ref{fig1}, the general solution of Eq.~(\ref{apeq17}) reads~\cite{Buyuk2012}
\be
\label{apeq19}
\tv(z,z')=4\pi\ell_{\rm B}\frac{h_+(z_<)h_-(z_>)+\Delta h_-(z_<)h_-(z_>)}{h'_+(z')h_-(z')-h'_-(z')h_+(z')},
\ee
where the functions $h_\pm(z)$ solve the homogeneous part of Eq.~(\ref{apeq17}),
\be
\label{apeq20}
\left[\partial_z^2-p^2(z)\right]h_{\pm}(z)=0.
\ee
Substituting the PB potential profile~(\ref{10}) into Eqs.~(\ref{apeq16}) and~(\ref{apeq18}), Eq.~(\ref{apeq20}) becomes
\be
\label{apeq21}
h''_\pm(z)-\left\{p^2+\frac{2\kappa^2}{\sinh^2\left[\kappa(z+z_0)\right]}\right\}h_\pm(z)=0,
\ee
with the parameter $p=\sqrt{k^2+\kappa^2}$ and the characteristic thickness of the interfacial cation layer $z_0=\ln(\gamma^{-1})/\kappa$. Eq.~(\ref{apeq21}) is solved by
\be\label{apeq22}
h_\pm(z)=e^{\pm pz}\left\{1\mp\frac{\kappa}{p}\coth\left[\kappa(z+z_0)\right]\right\}.
\ee
Using the solutions~(\ref{apeq22}), the Fourier-transformed Green's function~(\ref{apeq19}) can be simplified as
\be\label{apeq23}
\tv(z,z')=\frac{2\pi\ell_{\rm B}p}{k^2}\left[h_+(z_<)+\Delta h_-(z_<)\right]h_-(z_>)
\ee
where we introduced the delta function
\be\label{apeq24}
\Delta=\frac{\kappa^2\mathrm{csch}^2\left(\kappa z_0\right)+(p_{\rm b}- k)\left[p_{\rm b}-\kappa\coth\left(\kappa z_0\right)\right]}
{\kappa^2\mathrm{csch}^2\left(\kappa z_0\right)+(p_{\rm b}+ k)\left[p_{\rm b}+\kappa\coth\left(\kappa z_0\right)\right]},
\ee
and the coordinate variables $z_<=\mathrm{min}(z,z')$ and $z_>=\mathrm{max}(z,z')$. In the bulk limit $z\to\infty$ and $z'\to\infty$, the Green's function~(\ref{apeq23}) naturally tends to Eq.~(\ref{tvb}). Substituting now Eq.~(\ref{apeq23}) and~(\ref{tvb}) into Eqs.~(\ref{grsym}) and (\ref{ionself}), and passing to the dimensionless Fourier wave vector $u=k/\kappa$, after some algebra, the ionic self-energy finally follows as
\bea
\label{apeq25}
\delta v(\tz)&=&\Gamma_{\rm s}\int_1^{\infty}\frac{\mathrm{d}u}{u^2-1}\left\{-\mathrm{csch}^2\left(\tz+\tz_0\right)\right.\\
&&\hspace{2.4cm}\left.+\td\left[u+\mathrm{coth}\left(\tz+\tz_0\right)\right]^2e^{-2u\tz}\right\},\nonumber
\eea
with the delta function~(\ref{apeq24}) in dimensionless variables
\be
\label{apeq26}
\td=\frac{1+s\left(su-\sqrt{s^2+1}\right)\left(u-\sqrt{u^2-1}\right)}{1+s\left(su+\sqrt{s^2+1}\right)\left(u+\sqrt{u^2-1}\right)}.
\ee

\subsection{Computation of the \RP{$1\ell$} average potential correction}

We compute now the \RP{$1\ell$} correlation correction to the average electrostatic potential in Eq.~(\ref{potsup1}). This potential solves the differential equation
\be
\label{apeq27}
\partial_z^2\phi^{(1)}_{\rm m}(z)-\kappa_{\rm c}^2(z)\phi^{(1)}_{\rm m}(z)=-4\pi\ell_{\rm B}\delta\sigma(z),
\ee
with the charge excess function
\be
\label{apeq28}
\delta\sigma(z)=\rho_{\rm b}\sinh\left[\phi_0(\br)\right]\delta v(z).
\ee
We now note that the Fourier-transformed kernel $\tv(z,z')$ in Eq.~(\ref{apeq17}) is the Green's function of the differential equation~(\ref{apeq27}). Thus, using the definition of the Green's function
\be
\label{apeq29}
\int_{-\infty}^\infty\mathrm{d}z''\tv^{-1}(z,z'')\tv(z'',z')=\delta(z-z'),
\ee
the solution of Eq.~(\ref{apeq27}) can be expressed as
\be
\label{apeq30}
\phi^{(1)}_{\rm m}(z)=\int_0^\infty\mathrm{d}z'\tv(z,z';k=0)\delta\sigma(z').
\ee
Inserting now Eqs.~(\ref{apeq22})-(\ref{apeq25}) and~(\ref{apeq28}) into the integral of Eq.~(\ref{apeq30}), and carrying out the spatial integral, the \RP{$1\ell$} average potential correction follows as
\be
\label{apeq31}
\phi_{\rm m}^{(1)}(\tz)=\frac{\Gamma_{\rm s}}{4}\mathrm{csch}\left(\tz+\tz_0\right)\int_1^\infty\frac{\mathrm{d}u}{u^2-1}U(\tz),
\ee
with the auxiliary function
\bea
\label{apeq32}
U(\tz)&=&\frac{2+s^2}{s\sqrt{1+s^2}}-\td\left(\frac{1}{u}+2u+\frac{2+3s^2}{s\sqrt{1+s^2}}\right)\\
&&+\frac{\td}{u}e^{-2u\tz}+\left(\td\;e^{-2u\tz}-1\right)\coth\left(\tz+\tz_0\right).\nonumber
\eea
As noted at the beginning of Section~\ref{res}, Eqs.~(\ref{apeq25}) and~(\ref{apeq31}) show that the leading order correlation corrections to MF-level ion interactions and average membrane potential are proportional to the \RP{$1\ell$}-level salt coupling parameter $\Gamma_{\rm s}$.

\section{Derivation of the polymer grand potential components}
\label{1lpol}

In this Appendix, we explain the derivation of the polymer grand potential components in Eqs.~(\ref{eqs12})-(\ref{eqs15}) via the  inclusion of the monovalent salt correlations from the \RP{$1\ell$} electrostatic theory explained in Section~\ref{pots1l}. Below, these potentials will be derived for $0\leq\te\leq\pi/2$. Due to the mirror symmetry of the polymer grand potential with respect to the angle $\te=0$,  the grand potential can be evaluated for $\pi/2\leq\te\leq\pi$ by using the identity
\be
\label{apeq32II}
\Delta\Omega_\p(\tz_\p,\te)=\Delta\Omega_\p(\tz_\p,\pi-\te).
\ee

\subsection{Direct polymer-membrane charge coupling potential $\Omega_{\rm pm}$}

We derive here the \RP{$1\ell$}-level polymer-membrane interaction energy in Eq.~(\ref{eqs12}). Due to the linear superposition of the average MF potential and its \RP{$1\ell$} correction in Eq.~(\ref{potsup1}), the energy~(\ref{eqs12}) has a MF and \RP{$1\ell$} component,
\be
\label{apeq33}
\Omega_{\rm pm}(z_{\rm p},\theta_p)=\Omega^{(0)}_{\rm pm}(z_{\rm p},\theta_p)+\Omega^{(1)}_{\rm pm}(z_{\rm p},\theta_p).
\ee
The MF component of Eq.~(\ref{apeq33}) is given by Eq.~(\ref{11}). In order to derive the \RP{$1\ell$} component, we substitute into Eq.~(\ref{eqs12}) the average potential correction~(\ref{apeq31}). Carrying out the spatial integrals, after lengthy algebra, one obtains 
\be
\label{apeq34}
\beta\Omega^{(1)}_{\rm pm}(\tz_\p,\te)=-\frac{\Gamma_{\rm s}\tau}{2\kappa}\int_1^\infty\frac{\mathrm{d}u}{u^2-1}\frac{R(u)}{\cos\te},
\ee
where we introduced the auxiliary function
\begin{widetext}
\bea
\label{apeq35}
R(u)&=&S(u)\left[{\rm Arcth\left(\gamma e^{-\tz_-}\right)}-{\rm Arcth\left(\gamma e^{-\tz_+}\right)}\right]\\
&&+\frac{\gamma}{2}\td\left(1+u^{-1}\right)\left\{e^{-(2u+1)\tz_-}\Phi\left(\gamma^2e^{-2\tz_-},1,u+\frac{1}{2}\right)-e^{-(2u+1)\tz_+}\Phi\left(\gamma^2e^{-2\tz_+},1,u+\frac{1}{2}\right)\right\}\nonumber\\
&&+\td\left\{\gamma^{-2u}\left[\mb\left(\gamma^2e^{-2\tz_-},u+\frac{5}{2},-1\right)-\mb\left(\gamma^2e^{-2\tz_+},u+\frac{5}{2},-1\right)\right]\right.\nonumber\\
&&\hspace{8mm}\left.+\gamma^3\left[e^{-(2u+3)\tz_-}\Phi\left(\gamma^2e^{-2\tz_-},1,u+\frac{3}{2}\right)-e^{-(2u+3)\tz_+}\Phi\left(\gamma^2e^{-2\tz_+},1,u+\frac{3}{2}\right)\right]\right\}\nonumber\\
&&-\mb\left(\gamma^2e^{-2\tz_-},\frac{5}{2},-1\right)+\mb\left(\gamma^2e^{-2\tz_+},\frac{5}{2},-1\right)-\gamma^3\left[e^{-3\tz_-}\Phi\left(\gamma^2e^{-2\tz_-},1,\frac{3}{2}\right)-e^{-3\tz_+}\Phi\left(\gamma^2e^{-2\tz_+},1,\frac{3}{2}\right)\right].\nonumber
\eea
\end{widetext}
Eq.~(\ref{apeq35}) includes the function 
\be
\label{apeq36}
S(u)=\frac{2+s^2}{s\sqrt{1+s^2}}-\td\left(u^{-1}+2u+\frac{2+3s^2}{s\sqrt{1+s^2}}\right)-1,
\ee
the Lerch transcendent function
\be
\label{apeq37}
\Phi(x,n,a)=\sum_{i=0}^{\infty}\frac{x^i}{(i+a)^n},
\ee
and the incomplete Beta function 
\be
\label{apeq38}
\mb(x,a,b)=\int_0^x\md t\;t^{a-1}(1-t)^{b-1}.
\ee

\subsection{Polymer self-energy  $\Delta\Omega_{\rm pp}$}
\label{1lpp}

We compute now the polymer self-energy in Eq.~(\ref{eqs13}). In the corresponding formula, the double integral over the polymer charge position cannot be evaluated analytically. In order to simplify its numerical evaluation, we expand the homogeneous functions in Eq.~(\ref{apeq22}) in powers of the parameter $\gamma$ as
\be
\label{apeq39}
h_\pm(z)=\frac{\kappa}{p}\sum_{n\geq0}b_n^\mp e^{-v_n^\mp\tz},
\ee
with the expansion coefficients
\bea
\label{apeq40}
b_0^\pm=u\pm1;\hspace{2mm}b^\pm_{n>0}=\pm2\gamma^{2n};\hspace{2mm}v_n^\pm=2n\pm u.
\eea
Evaluating now the integrals in  Eq.~(\ref{eqs13}) with the Green's function~(\ref{apeq23}) and Eq.~(\ref{apeq39}), after long algebra, the \RP{$1\ell$} polymer self-energy renormalized by its bulk limit
\be\label{netpp}
\Delta\Omega_{\rm pp}(z_\p,\te)=\Omega_{\rm pp}(z_\p,\te)-\Omega_{\rm pp}(z_\p\to\infty,\te)
\ee
takes the form
\be
\label{apeq41}
\beta\Delta\Omega_{\rm pp}(\tz_\p,\te)=\frac{\Gamma_{\rm s}\tau^2}{2\kappa^2}\zeta_{\rm pp}(\tz_\p,\te).
\ee
In Eq.~(\ref{apeq41}), we introduced the dimensionless self-energy 
\begin{widetext}

\be
\label{apeq42}
\zeta_{\rm pp}(\tz_\p,\te)=\int_0^{2\pi}\frac{\md\phi_k}{2\pi}\int_1^\infty\frac{\mathrm{d}u}{u^2-1}\left\{F(u)-\frac{u^2-1}{\left(u^2\cos^2\te+q^2\right)^2}J(u)
+\td\left[G_{{\rm r}+}^2(u)+G_{{\rm c}+}^2(u)\right]\right\},
\ee
with the auxiliary functions

\bea
\label{apeq43}
&&G_{{\rm r}\pm}(u)=2\sum_{n=0}^\infty\frac{b_n^+}{t{_n^+}^2+q^2}\left\{t_n^+\sinh\left(\frac{t_n^+\tL}{2}\right)\cos\left(\frac{q\tL}{2}\right)+q\cosh\left(\frac{t_n^+\tL}{2}\right)\sin\left(\frac{q\tL}{2}\right)\right\}e^{-v_n^+\tz_\p},\\
\label{apeq44}
&&G_{{\rm c}\pm}(u)=2\sum_{n=0}^\infty\frac{b_n^+}{t{_n^+}^2+q^2}\left\{q\sinh\left(\frac{t_n^+\tL}{2}\right)\cos\left(\frac{q\tL}{2}\right)-t_n^+\cosh\left(\frac{t_n^+\tL}{2}\right)\sin\left(\frac{q\tL}{2}\right)\right\}e^{-v_n^+\tz_\p},\\
\label{apeq45}
&&F(u)=2\sum_{n,m\geq0}b_n^+b_m^-\;\left\{\frac{e^{-(t_n^+-t_m^-)\frac{\tL}{2}}}{\left(t{_n^+}^2+q^2\right)\left(t{_m^-}^2+q^2\right)}\left[-\left(t_n^+t_m^-+q^2\right)\cos(q\tL)+q(t_m^--t_n^+)\sin(q\tL)\right]\right.\nonumber\\
&&\left.\hspace{3.5cm}+\frac{t_n^+e^{(t_n^++t_m^-)\frac{\tL}{2}}}{\left(t{_n^+}^2+q^2\right)(t_n^++t_m^-)}+\frac{t_m^-e^{-(t_n^++t_m^-)\frac{\tL}{2}}}{\left(t{_m^-}^2+q^2\right)(t_n^++t_m^-)}\right\}\;e^{-(v_n^++v_m^-)\tz_\p},\\
\label{apeq45II}
&&J(u)=2u\tL\ct\left[u^2\cos^2\te+q^2\right]-4uq\ct e^{-u\tL\ct}\sin(q\tL)-2\left(u^2\cos^2\te-q^2\right)\left[1-e^{-u\tL\ct}\cos(q\tL)\right].\nonumber\\
\eea
\end{widetext}
Eqs.~(\ref{apeq43})-(\ref{apeq45II}) includes the additional coefficients 
\bea\label{apeq45III}
t_n^{\pm}&=&v_n^{\pm}\ct\\
\label{apeq45IV}
q&=&\sqrt{u^2-1}\sin\te\cos\phi_k.  
\eea
We also note that in Eq.~(\ref{apeq42}), the term proportional to the function $J(u)$ substracts the bulk self-energy in the reservoir at $z_\p\to\infty$. This bulk energy was obtained from Eq.~(\ref{eqs13}) by replacing the Fourier-transformed Green's function by its bulk limit~(\ref{tvb}).

\subsection{Polymer-counterion interaction energy  $\Delta\Omega_{\rm pc}$}
\label{1lpc}

The polymer-counterion coupling energy~(\ref{eqs14}) renormalized by its bulk limit is
\be\label{netpc}
\Delta\Omega_{\rm pc}(z_\p,\te)=\Omega_{\rm pp}(z_\p,\te)-\Omega_{\rm pc,b},
\ee
with the bulk energy $\Omega_{\rm pc,b}=\Omega_{\rm pc}(z_\p\to\infty,\te)$ in the reservoir region $z_\p\to\infty$. We compute first the bulk part of Eq.~(\ref{netpc}). To this end, we will use the three dimensional Fourier expansion of the bulk Green's function~(\ref{18})
\be
\label{apeq7}
v_{\rm b}(\br-\br')=\int\frac{\mathrm{d}^3\bq}{(2\pi)^3}u_{\rm b}(q)e^{i\bq\cdot(\br-\br')},
\ee
with $u_{\rm b}(q)=4\pi\ell_{\rm B}/\left(\kappa^2+q^2\right)$. Substituting the Fourier expansion~(\ref{apeq7}) into Eq.~(\ref{eqs8}), one finds
\be
\label{apeq8}
\beta\Omega_{\rm pc,b}=q_\C\rho_{\rm bc}\int\mathrm{d}\br\sigma_\p(\br)\int\frac{\mathrm{d}^3\bq}{(2\pi)^3}u_{\rm b}(q)\int\mathrm{d}\br_\C e^{i\bq\cdot(\br-\br_\C)}.
\ee
The first integral yields the net polymer charge $-\tau L$ while the integral over the counterion position $\br_\C$ generates the delta function $\left(2\pi\right)^3\delta^3(\bq)$. Finally, Eq.~(\ref{apeq8}) simplifies to the expression
\be
\label{apeq9}
\beta\Omega_{\rm pc,b}=- \frac{4\pi\ell_{\rm B}\rho_{\rm bc}\qc}{\kappa^2}L\tau
\ee
corresponding to Eq.~(\ref{pcb}) in the main text.

In order to derive the net polymer-counterion interaction energy~(\ref{netpc}), we insert into Eq.~(\ref{eqs14}) the Fourier-transformed Green's function~(\ref{apeq23}) together with the expansion in Eq.~(\ref{apeq39}). Carrying out the spatial integral over the variable $l$, after lengthy algebra, one obtains
\be
\label{apeq9II}
\Delta\Omega_{\rm pc}(z_\p,\te)=\frac{4\pi\ell_{\rm B}\rho_{\rm bc}q_\C }{\kappa^2}L\tau\left[1-F(z_\p,\te;u=1)\right],
\ee
where we introduced the auxiliary function
\begin{widetext}
\bea
\label{apeq9III}
F(u)&=&\frac{1}{2\tL(u^2-1)}\left\{G_{{\rm r}+}(u)\int_0^{\zm}\mathrm{d}\tz k_\C(\tz)\left[h_+(\tz)+\td h_-(\tz)\right]+\int_{\zm}^{\zp}\mathrm{d}\tz k_\C(\tz)V_{\rm r}(\tz;u)\right.\nonumber\\
&&\hspace{2cm}\left.+\left[G_{{\rm r}-}(u)+\td G_{{\rm r}+}(u)\right]\int_{\zp}^\infty\mathrm{d}\tz k_\C(\tz)h_-(\tz)\right\}.
\eea
Eq.~(\ref{apeq9III}) contains the dimensionless counterion density obtained from Eq.~(\ref{eqs11}) as $k_\C(z)=\exp\left[-q_{\rm c}^2\delta v(z)/2-q_{\rm c}\phi_{\rm m}(z)\right]$, and the additional functions
\bea
\label{apeq9V}
V_{\rm r}(\tz;u)&=&h_-(\tz)\sum_{n=0}^\infty\frac{b_n^-}{t{_n^-}^2+q^2}\left\{\left(-t_n^-\cos\left[q\lt\right]+q\sin\left[q\lt\right]\right)e^{-v_n^-\tz}+\left(t_n^-\cos\left[q\tL/2\right]+q\sin\left[q\tL/2\right]\right)e^{-v_n^-\zm}\right\}\nonumber\\
&&+h_+(\tz)\sum_{n=0}^\infty\frac{b_n^+}{t{_n^+}^2+q^2}\left\{\left(t_n^+\cos\left[q\lt\right]-q\sin\left[q\lt\right]\right)e^{-v_n^+\tz}+\left(-t_n^+\cos\left[q\tL/2\right]+q\sin\left[q\tL/2\right]\right)e^{-v_n^+\zp}\right\}\nonumber\\
&&+\td G_{{\rm r}+}(u)h_-(\tz),\\
\label{apeq9VI}
V_{\rm c}(\tz;u)&=&h_-(\tz)\sum_{n=0}^\infty\frac{b_n^-}{t{_n^-}^2+q^2}\left\{-\left(t_n^-\sin\left[q\lt\right]+q\cos\left[q\lt\right]\right)e^{-v_n^-\tz}+\left(-t_n^-\sin\left[q\tL/2\right]+q\cos\left[q\tL/2\right]\right)e^{-v_n^-\zm}\right\}\nonumber\\
&&+h_+(\tz)\sum_{n=0}^\infty\frac{b_n^+}{t{_n^+}^2+q^2}\left\{\left(t_n^+\sin\left[q\lt\right]+q\cos\left[q\lt\right]\right)e^{-v_n^+\tz}-\left(t_n^+\sin\left[q\tL/2\right]+q\cos\left[q\tL/2\right]\right)e^{-v_n^+\zp}\right\}\nonumber\\
&&+\td G_{{\rm c}+}(u)h_-(\tz),
\eea
\end{widetext}
where $\lt=(\tz-\tz_\p)/\ct$. The function defined in Eq.~(\ref{apeq9VI}) will be used in Appendix~\ref{1lpcp}. We finally note that the spatial integrals in Eq.~(\ref{apeq9III}) should be evaluated numerically.

\subsection{Screening energy of the polymer self-interaction $\Delta\Omega_{\rm pcp}$}
\label{1lpcp}

We finally compute the screening energy of the polymer self-interaction in Eq.~(\ref{eqs15}) renormalized by its bulk value,
\be\label{netpcp}
\Delta\Omega_{\rm pcp}(z_\p,\te)=\Omega_{\rm pcp}(z_\p,\te)-\Omega_{\rm pcp,b},
\ee
where we defined the bulk energy $\Omega_{\rm pcp,b}=\Omega_{\rm pcp}(z_\p\to\infty,\te)$ in the reservoir $z_\p\to\infty$. To derive first the bulk component of Eq.~(\ref{netpcp}), we substitute the Fourier expanded bulk Green's function~(\ref{apeq7}) into Eq.~(\ref{eqs9}). This yields
\begin{widetext}
\be
\label{apeq10}
\beta\Omega_{\rm pcp,b}=-\frac{q_\C^2\rho_{\rm bc}\tau^2}{2}\int_{-L/2}^{L/2}\mathrm{d}l_1\int_{-L/2}^{L/2}\mathrm{d}l_2\int_0^\infty\frac{\mathrm{d}qq^2}{(2\pi)^3}u_{\rm b}^2(q)J(q,l_1,l_2),
\ee
with the auxiliary integral
\be
\label{apeq11}
J(q,l_1,l_2)=\int_0^{2\pi}\mathrm{d}\varphi_q\int_0^\pi\mathrm{d}\theta_q\sin\theta_q\;e^{i\bq\cdot\left[\br(l_1)-\br(l_2)\right]},
\ee
where $(\varphi_q,\theta_q)$ are the spherical angles in the reciprocal Fourier space. In Eq.~(\ref{apeq11}), we employed as well the parametric description of the position vector on the polymer, $\br(l)=x(l)\hat{u}_x+y(l)\hat{u}_y+z(l)\hat{u}_z$. Substituting the corotating coordinates in Eqs.~(\ref{c1})-(\ref{c3}) into Eq.~(\ref{apeq11}), one gets
\be
\label{apeq12}
J(q,l_1,l_2)=\int_0^{2\pi}\mathrm{d}\varphi_q\int_0^\pi\mathrm{d}\theta_q\sin\theta_q\;e^{i\bq\cdot\left(\bl_1-\bl_2\right)}=\int_0^{2\pi}\mathrm{d}\varphi_q\int_0^\pi\mathrm{d}\theta_q\sin\theta_q\;e^{iq\left|l_1-l_2\right|\cos\theta_q}
\ee
Carrying out the angular integrals and substituting the result into Eq.~(\ref{apeq10}), one obtains
\be
\label{apeq13}
\beta\Omega_{\rm pcp,b}=-4\rho_{\rm bc}q_\C^2\ell_{\rm B}^2\tau^2\int_{-L/2}^{L/2}\mathrm{d}l_1\int_{-L/2}^{L/2}\mathrm{d}l_2\int_0^\infty\frac{\mathrm{d}qq^2}{\left(\kappa^2+q^2\right)^2}\frac{\sin\left[q\left|l_1-l_2\right|\right]}{q\left|l_1-l_2\right|}.
\ee
Evaluating first the Fourier integral, and then carrying out the integrations over the coordinates $l_{1,2}$ of the polymer charges, one finally gets 
\be
\label{apeq14}
\beta\Omega_{\rm pcp,b}=-2\pi\left(\ell_{\rm B}\tau\right)^2\frac{\rho_{\rm bc}q_\C^2}{\kappa^3}\left(\kappa L+e^{-\kappa L}-1\right).
\ee

The net screening energy~(\ref{netpcp}) is obtained by substituting into Eq.~(\ref{eqs15}) the Green's function~(\ref{apeq23}) and the expansion in Eq.~(\ref{apeq39}). Evaluating the integral over the coordinate $l$ of the polymer charges, after rather long algebra, one obtains
\bea\label{apeq46}
\Delta\Omega_{\rm pcp}(z_\p,\te)&=&2\pi\left(\ell_{\rm B}\tau\right)^2\frac{\rho_{\rm bc}\qc^2}{\kappa^3}\left\{\tL+e^{-\tL}-1-\int_0^{2\pi}\frac{\md\phi_k}{4\pi}\int_1^\infty\frac{\mathrm{d}u}{(u^2-1)^2}Y(u)\right\},
\eea
with the auxiliary function
\bea
\label{apeq47}
Y(u)&=&\left[G^2_{{\rm r}+}(u)+G^2_{{\rm c}+}(u)\right]\int_0^{\zm}\mathrm{d}\tz k_\C(\tz)\left[h_+(\tz)+\td h_-(\tz)\right]^2+\int_{\zm}^{\zp}\mathrm{d}\tz k_\C(\tz)\left[V_{\rm r}^2(\tz;u)+V_{\rm c}^2(\tz;u)\right]\\
&&+\left\{\left[G_{{\rm r}-}(u)+\td G_{{\rm r}+}(u)\right]^2+\left[G_{{\rm c}-}(u)+\td G_{{\rm c}+}(u)\right]^2\right\}\int_{\zp}^\infty\mathrm{d}\tz k_\C(\tz)h^2_-(\tz).
\eea

\end{widetext}

\end{document}